\documentclass{aa}  
\usepackage{subcaption}
\usepackage{placeins}
\usepackage{hyperref}
\usepackage{float}
\usepackage{natbib}
\usepackage[utf8]{inputenc}
\usepackage[T1]{fontenc}
\usepackage{arydshln}

\begin{document}
   \title{4MOST ChANGES:  Catalog of high-redshift quasar candidates (4.5 < $z$ < 7) selected with SED fitting}
  \titlerunning{Catalog of high-redshift quasar candidates}
   \authorrunning{Mkrtchyan et al.}
  \author{Tatevik Mkrtchyan\raisebox{-0.7ex}{\includegraphics[height=1.1em]{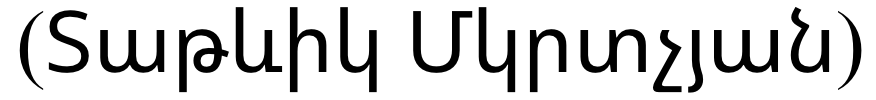}\inst{1}}
          \and
        Chiara Mazzucchelli\inst{1}
        \and
        Roberto J. Assef\inst{1}
        \and
        Matthew J. Temple\inst{2}
        \and
        Alejandra Rojas-Lilayú\inst{3}
        \and 
        Franz E. Bauer\inst{4}
        \and 
        Victoria Toptun\inst{5}
        \and 
        J. A. Acevedo Barroso\inst{14}
        \and 
        Silvia Belladitta\inst{7,8}
        \and 
        Emanuele Paolo Farina\inst{8,9}
        \and 
        L. N. Martínez-Ramírez\inst{6,10,11}
        \and 
        Giulia Papini\inst{8,12}
        \and 
        Sarath Satheesh-Sheeba\inst{13}
        \and 
        Daniel Stern\inst{14}
        \and 
        Anuraag Upadhyayula\inst{15}
}
        
\institute{Instituto de Estudios Astrofísicos, Universidad Diego Portales, Av. Ejército Libertador 441, Santiago, Chile
          \and 
          Centre for Extragalactic Astronomy, Department of Physics, Durham University, South Road, Durham DH1 3LE, United Kingdom
          \and
          Universidad Técnica Federico Santa María, Av. Vicuña Mackenna 3939, 8940897 San Joaquín, Santiago, Chile
          \and
          Instituto de Alta Investigaci{\'{o}}n, Universidad de Tarapac{\'{a}}, Casilla 7D, Arica, Chile
          \and
          European Southern Observatory, Karl Schwarzschildstrasse 2, 85748, Garching bei München, Germany
          \and
          Instituto de Astrofísica, Facultad de Física, Pontificia Universidad Católica de Chile Av. Vicuña Mackenna 4860, 7820436 Macul, Santiago, Chile
          \and
          Max-Planck-Institut für Astronomie, Königstuhl 17, D-69117, Heidelberg, Germany
          \and
          INAF, Osservatorio di Astrofisica e Scienza dello Spazio (OAS), Via Gobetti 93/3, 40129 Bologna, Italy
          \and
          International Gemini Observatory/NSF NOIRLab, 670 N A’ohoku Place, Hilo, Hawai'i 96720, USA
          \and
          Fakultät für Physik und Astronomie, Universität Heidelberg, Im Neuenheimer Feld 226, 69120 Heidelberg, Germany
          \and
          Millennium Institute of Astrophysics (MAS), Nuncio Monseñor Sótero Sanz 100, Providencia, Santiago, Chile
          \and
          Dipartimento di Fisica e Astronomia "Augusto Righi", Alma Mater Studiorum - Università di Bologna, via Gobetti 93/2, 40129 Bologna, Italy
          \and
          Instituto de Astrofísica, Facultad de Ciencias Exactas, Universidad Andrés Bello, Fernández Concha 700, 7591538 Las Condes, Santiago, Chile 
          \and
          Jet Propulsion Laboratory, California Institute of Technology, 4800 Oak Grove Drive, Pasadena, CA 91109, USA
          \and
          University of Missouri–Kansas City, Kansas City, MO 64110, USA
          }
 
  \abstract
   {The identification of high-redshift quasars ($z > 4.5$) is critical for studying the early Universe, supermassive black hole growth, and cosmic reionization. Most known high-redshift quasars are located in the northern hemisphere, leaving the southern sky largely unexplored.}
   {As part of the 4-meter Multi-Object Spectroscopic Telescope (4MOST) and Chilean AGN/Galaxy Extragalactic Survey (ChANGES) S1604 survey, we aim to create a large catalog of high-redshift quasar candidates in the southern hemisphere using multiwavelength photometry and Spectral Energy Distribution (SED) fitting, with the goal of spectroscopic follow-up with 4MOST.}
   {We construct a multi-band photometric catalog by combining optical data from the Dark Energy Camera Local Volume Exploration Survey (DELVE DR2) and Dark Energy Camera Legacy Survey (DECaLS DR10), near-infrared data from the VISTA Hemisphere Survey (VHS DR5) with an additional field of the VISTA Kilo-degree Infrared Galaxy Public Survey (VIKING), mid-infrared data from the Wide-field Infrared Survey Explorer (AllWISE) and optical astrometry from Gaia DR3. After applying morphological and color-based cuts to remove contaminants such as brown dwarfs and red galaxies, we perform a custom-made SED fitting using quasar and brown dwarf templates. Statistical outputs including $\chi^2$, Bayesian Information Criterion $(BIC)$, and $F_{\text{test}}$ are used to rank and select candidates.}
   {Our final catalog contains 6104 high-redshift quasar candidates within the redshift range of $4.5 < z < 7$. These sources have detections in 7 or more photometric bands and satisfy our SED-based statistical selection criteria (e.g. $BIC > 0$ and $F_{\text{test}} > 10$). Initial spectroscopic validation using the New Technology Telescope (NTT) with the ESO Faint Object Spectrograph and Camera v.2 (EFOSC2) and the Palomar Observatory Hale Telescope with the Next Generation Palomar Spectrograph (NGPS) confirmed 3 high-redshift quasars at $z > 5$ out of 6 observed candidates.}
   {}

   \date{Accepted 9 April 2026}
   \maketitle

\clearpage
\section{Introduction}
\paragraph{} 
Quasars are the most luminous type of Active Galactic Nuclei (AGN) powered by the accretion of material onto supermassive black holes (SMBHs) located at the centers of massive galaxies \citep[e.g.,][]{Magorrian_1998, Marconi_2003}. These powerful sources emit across the entire electromagnetic spectrum. They can be studied at large cosmological look-back times, with spectroscopically confirmed quasars now reaching redshifts of $z \sim 7.6$ \citep[e.g.,][]{Banados_2018, Yang_2020a, Wang_2021b}. Recent observations have also identified AGN candidates at even higher redshifts of approximately $z \sim 10$ (about half a billion years after the Big Bang; e.g., \citealt{Bogdan_2024}; \citealt{Natarajan_2023}).
Additionally, high-redshift quasars provide valuable information on the chemical composition and metal enrichment of the intergalactic medium (IGM; e.g., \citealt{Becker_2015}; \citealt{Bosman_2022}; \citealt{Davies_2023}) and highlight relatively dense environments (e.g., \citealt{Mignoli_2020}; \citealt{Pudoka_2024}; \citealt{Lambert_2024}; \citealt{Champagne_2025a}; \citealt{Champagne_2025b}).

\noindent The discovery of the first quasars at $z > 4$ in the 1980s marked a significant milestone in our understanding of the distant Universe \citep[e.g.,][]{Warren_1987}. The exploration of high-redshift quasars further accelerated with the advent of the Sloan Digital Sky Survey \citep[SDSS;][]{York_2000}, which led to the discovery of the first quasars at $z > 5$ \citep[e.g.,][]{Fan_2001}. Since then, systematic wide-field surveys such as Wide-field Infrared Survey Explorer \citep[WISE;][]{Wright_2010}, the UKIRT Infrared Deep Sky Survey \citep[UKIDSS;][]{Lawrence_2007}, Gaia \citep{Brown_2016} and the Panoramic Survey Telescope and Rapid Response System \citep[Pan-STARRS;][]{Chambers_2016}, along with deep, multiwavelength observations of smaller fields like COSMOS and the Chandra Deep Fields, have significantly advanced the identification and characterization of AGN and quasars beyond the local Universe \citep{Brandt_2001, Giacconi,Lehmer_2005,Salvato_2009,Marchesi_2016}. To date, more than 500 quasars have been identified at $z > 5.6$, the majority located in the northern hemisphere \citep{Fan_2023, Yang_2024, Belladitta_2025}. These discoveries confirmed the existence of SMBHs with masses up to a few billion solar masses ($M_{\odot}$) within the first billion years after the Big Bang, with rest-frame UV/optical spectra remarkably similar to those of luminous low-redshift quasars \citep[e.g.,][]{Mazzucchelli_2023,Farina_2022}. The field has been recently revolutionized by the James Webb Space Telescope (JWST), which has revealed new populations of faint, obscured high-redshift AGN candidates that were previously missed by traditional selection methods, specifically Little Red Dots (LRDs), with number densities an order of magnitude higher than expected \citep[e.g.,][]{Matthee_2024,Kocevski_2024}.

Most high-redshift quasars have been identified using the Lyman-break (or dropout) technique \citep{Warren_1987, Steidel_1996}. This method capitalizes on the absorption of flux below the Ly$\alpha$ emission line (at a rest-frame wavelength of 1215 Å) due to both the Lyman break and the Ly$\alpha$ forest from intervening absorption features, which together produce very red broadband colors. Early work on photometric quasar identification established color-based selection criteria \citep{Richards_2001, Richards_2004} that have been continuously refined over two decades \citep{Richards_2015}. The color cut method is straightforward to implement and achieves high completeness, even for quasars with atypical spectra \citep{Fan_2023}. However, this method often suffers from significant contamination due to the similar optical and near-infrared (NIR) colors of other astrophysical sources. Multiple stellar populations have similar colors to high-redshift quasars, including cool brown dwarfs with spectral types M, L, and T, low-mass stars, and compact early-type galaxies \citep[e.g.,][]{Hewett_2006, Findlay_2012,  Venemans_2015, Banados_2016, Gloudemans_2022, Belladitta_2025, Ighina_2025}. Among these contaminants, cool brown dwarfs represent the dominant source of confusion in photometric surveys, outnumbering $z\sim7$ quasars by 2–4 orders of magnitude in deep surveys \citep{Barnett_2019}, making them the primary challenge for color-based selection of high-redshift quasars. 

To address these issues, several studies \cite[e.g.,][]{Hatziminaoglou_2004, Mortlock_2012, Matsuoka_2016, Reed_2017} applied Spectral Energy Distribution (SED) fitting following the initial color cuts, calculating reduced $\chi^2$ values by comparing observed SEDs to models of quasars and potential contaminants, including various M, L, T dwarf types and early-type galaxies at different redshifts. These SED-fitting approaches have been refined with multiwavelength data over the years \citep{Messias_2012, Findlay_2012} and extended to higher redshifts \citep{Nakoneczny_2021}, including recent searches for dust-reddened quasars at $z > 6$ \citep{Iwamoto_2025}.More recently, the use of machine learning techniques has facilitated the discovery of new high-redshift quasars, even in datasets that have previously undergone extensive searches \cite[e.g.,][]{Wenzl_2021, Yang_2024, Byrne_2024, Calderone_2024}.
  
In addition to northern hemisphere efforts such as SDSS \citep{York_2000} and Pan-STARRS \citep[PS1;][]{Chambers_2016} that have enabled a drastic increase in the number of quasars discovered at $z > 4$ \citep[e.g.,][]{Shen_2011, Banados_2016, Banados_2023, Jiang_2016, Wang_2016, Gloudemans_2022, Belladitta_2019, Belladitta_2023}, several recent large-area surveys have expanded high-redshift quasar searches into the southern sky, such as the Dark Energy Survey \citep[DES;][]{Flaugher_2005,Abbott_2018} the DESI Legacy Imaging Surveys \citep[LS;][]{Dey_2019}, and SkyMapper \citep{Keller_2007}, which have enabled the discovery of high-redshift quasars in the southern hemisphere \citep[e.g.,][]{Reed_2017,Pons_2019, Belladitta_2019,Wolf_2020,Onken_2022,Yang_2023,Ighina_2023, Ighina_2025}. However, the southern hemisphere still lacks the extensive spectroscopic coverage and time-domain infrastructure available in the north. This gap presents both a challenge and an opportunity for the next generation of extragalactic surveys \citep[e.g.,][]{Jong_2019, Bauer_2023}.

Among the various astronomical facilities, the 4-meter Multi-Object Spectroscopic Telescope (4MOST) installed on the Visible and Infrared Survey Telescope for Astronomy (VISTA) at Paranal Observatory stands out as a key asset in the southern hemisphere, covering the sky in the declination range $-70^\circ < \text{dec} < 5^\circ$ in the coming years \citep{Guiglion_2019}. 4MOST can observe approximately 2,400 targets simultaneously across a wide field of view of 4.2 square degrees. Over its initial five-year survey, it is expected to obtain spectra for over 20 million sources at a spectral resolution of $R \sim 6500$, and more than 3 million spectra at $R \sim 20,000$ \citep{Jong_2019}. 

The 4MOST Chilean AGN/Galaxy Evolution Survey \citep[ChANGES;][]{Bauer_2023} is one of 15 community surveys approved for 4MOST, allocated approximately 8.23\% of the low-resolution spectrograph (LRS) fiber hours. ChANGES aims to obtain spectroscopy for $\sim$2.3 million AGN candidates spanning $\sim$18,000 square degrees, primarily selected through optical variability (using ZTF, La Silla QUEST, and Gaia data) and optical/NIR/mid-infrared (MIR) SED fitting. The survey's primary science goals include: (1) improving population statistics for moderate-luminosity and lower-mass black hole AGN out to $z \sim 1$; (2) expanding the known AGN sample available to facilities such as the Vera C. Rubin Observatory's Legacy Survey of Space and Time \citep[LSST;][]{Ivezic_2019}, Euclid, and ESO's Extremely Large Telescope; (3) generating a massive training set for LSST/Euclid classification and photometric redshift calibration; (4) studying AGN spectral variability through multi-epoch observations of $\sim$200,000 variable-selected AGN; and (5) investigating extreme variability events, tidal disruption events, and lensed AGN through dedicated target-of-opportunity observations.

Among these goals, ChANGES dedicates effort to identifying high-redshift ($4.5 < z < 7$) quasar candidates in the southern hemisphere, particularly at the southernmost declinations ($\text{dec} < -30^\circ$), which remain poorly explored. Our work aims to provide a catalog of high-redshift quasar candidates selected using a custom-made SED fitting code and applying several criteria outlined in the following sections. The structure of the paper is organized as follows: In \hyperref[sec:data]{Section~2}, we describe the data used in this work. \hyperref[sec:method]{Section~3} explains the method of photometric selection and provides justification for the criteria applied. \hyperref[sec:results]{Section~4} presents the output of our catalog with the results obtained and the preliminary spectroscopic follow-up. Finally, we summarize our findings in \hyperref[sec:summary]{Section~5}. We used a flat $\Lambda$ cold dark matter ($\Lambda$CDM) cosmology with $H_0 = 70$ km s$^{-1}$ Mpc$^{-1}$, $\Omega_{\rm m} = 0.30$, and $\Omega_{\Lambda} = 0.70$.

\section{Photometric Data}
\label{sec:data}
\paragraph{} 
To ensure consistency across multiple ChANGES science cases, a single uniform photometric catalog is constructed, which includes $\sim 4.2\times10^8$ objects from the southern sky ($dec < 5^\circ$). The catalog is created by cross-matching optical $griz$ photometry from the second public data release (DR2) of the Dark Energy Camera Local Volume Exploration Survey \citep[DELVE;][]{Drlica_Wagner_2022}, NIR $YJHK_s$ photometry from the fifth data release (DR5) of the VISTA Hemisphere Survey (VHS) with an additional field of the VISTA Kilo-degree Infrared Galaxy Public Survey \citep[VIKING;][]{Edge_2013}, and MIR $W1$ and $W2$ photometry from the CatWISE 2020 catalog \citep{Marocco_2020}. The primary source detections are from DELVE DR2, with additional photometry cross-matched from VHS/VIKING and CatWISE when available, but not required. NIR gaps in the southern sky coverage are supplemented by VIKING, which covers approximately $0^\circ \lesssim RA \lesssim 50^\circ$ and $-35^\circ \lesssim dec \lesssim -10^\circ$, addressing a major gap in VHS coverage. Additional gaps in VHS coverage, including several horizontal stripes at various declinations (most prominently at $\mathrm{dec} \sim -5^\circ$, $-15^\circ$, $-30^\circ$, and $-40^\circ$) and scattered regions throughout the footprint, cannot be filled with NIR data. In these regions where NIR data from VHS/VIKING are unavailable, the catalog contains only optical and MIR photometry (DELVE+CatWISE). The different photometric depths between VHS (5$\sigma$ limiting magnitudes: $J \sim 20.6$, $K_s \sim 18.5$; \citealt{McMahon+2013, McMahon_2020}) and VIKING ($J \sim 21.2$, $K_s \sim 19.8$; \citealt{Edge_2013}) may introduce systematic biases in candidate selection across different sky regions. Duplicate objects are removed using a $1\arcsec$ matching radius, consistent with the $1.4\arcsec$ diameter of the 4MOST fibers. Furthermore, this catalog incorporates proper motion information and object-type flags from Gaia DR3 \citep{Vallenari_2022}.
In a following step of our selection, we additionally use magnitudes from the tenth data release (DR10) of the Dark Energy Camera Legacy Survey (DECaLS;  \citealt{Dey_2019}\footnote{https://www.legacysurvey.org/dr10/description/}), which are not included in our initial catalog as this catalog was built to optimize sky coverage and consistency across all ChANGES science cases, hence DELVE DR2 was chosen instead for the optical coverage (Assef in prep, Bauer in prep). We also use the Milky Way dust reddening map, from Schlegel, Finkbeiner, \& Davis \citep[SFD;][]{Schlegel_1998} dust maps to obtain dust extinction values $E(B-V)$. The catalog is divided into sections of $15^\circ$ by right ascension (RA), each containing around 8 million to 20 million entries.   

The Dark Energy Camera (DECam) is a 570-megapixel instrument mounted on the 4-meter Victor M. Blanco Telescope at Cerro Tololo in Chile. Since its commissioning, DECam has served the DES, the DECaLS, and other community programs, capturing much of the southern sky. DELVE DR2 includes imaging from DELVE, DES and DECaLS with a median $5\sigma$ point-source depth of $g = 24.3$, $r = 23.9$, $i = 23.5$, and $z = 22.8$ magnitudes \citep{Drlica_Wagner_2022}, and covers nearly the entire Southern hemisphere. In contrast, DECaLS DR10 has new $i$-band imaging plus $grz$. It incorporates DECaLS observations from 2014 to 2019 and additional DECam data from DES, DELVE, and DeROSITA surveys. DECaLS reaches depths of $g=24.7$, $r=23.9$, $z=23.0$, referring to the target $5\sigma$ point-source depth, providing deeper $g$-band photometry that serves as a helpful alternative selection criterion for removing potential contaminants. However, DECaLS DR10 does not provide full sky coverage in the 4MOST accessible area, with significant gaps particularly in the $i$ and $z$ bands, which is why DELVE DR2 serves as our primary photometric source.

The VISTA Hemisphere Survey DR5 spans the entire southern celestial hemisphere ($dec < 0^\circ$), covering 20,000 $deg^2$ and reaching depths 30 times fainter than those achieved by 2MASS/DENIS. This survey provides data for nearly $\sim1.4$ billion detections, offering measurements across 4 NIR broadband filters such as Y, J, H, and Ks with AB depths of 21.1, 20.8, 20.5, 20.0 (5$\sigma$), respectively \citep{Pons_2019}. Similarly, VIKING is covering 1500 $deg^2$ across two regions of the extragalactic sky using VISTA in the $z$, $Y$, $J$, $H$, and $K_s$ bands to an AB depth of 23.1, 22.3, 22.1, 21.5, and 21.2, respectively, with the $5\sigma$ point-source depth \citep{Edge_2013}. Lastly, we used the CatWISE 2020 Catalog, which combines 2-band (3.4, 4.6) fluxes, positional data, apparent motion measurements, and flux variability statistics \citep{Wright_2010, Marocco_2020}. 

In summary, our initial photometric catalog covers the full southern sky ($dec < 5^\circ$) and contains $\sim 4.2 \times 10^8$ objects with up to 10 bands of DECam $griz$, VISTA $YJHK_s$, and WISE $W1/2$, spanning a wavelength range of $\sim$4800--46000\,\AA\ (optical to MIR). For magnitude measurements, we use Kron-like aperture photometry (\texttt{mag\_auto}) for DELVE optical data, Petrosian photometry (\texttt{petromag}) for VHS/VIKING NIR data, and profile-fitting photometry (\texttt{mpro}) for CatWISE MIR data. All magnitudes are in the AB system unless otherwise stated. In addition to photometric data, some value-added (non-photometric) columns are used in the selection process. These are 3 Gaia-based parameters: proper motion if available and its uncertainties in RA and Dec, the $extended\_class\_z\_delve$ flag, and $E(B-V)$ values derived from SFD.

The full catalog is divided into sections of 15 degrees by RA, with each segment containing around 8 million to 20 million entries, including 3 chunks with the gaps in VHS coverages.  It is publicly available\footnote{\url{https://gal-04.voxastro.org/allsky_2022/}}.

\section{Methodology} 
\label{sec:method}

\begin{figure}[!htbp]
     \centering
\includegraphics[width=\hsize,height=23 cm, keepaspectratio]{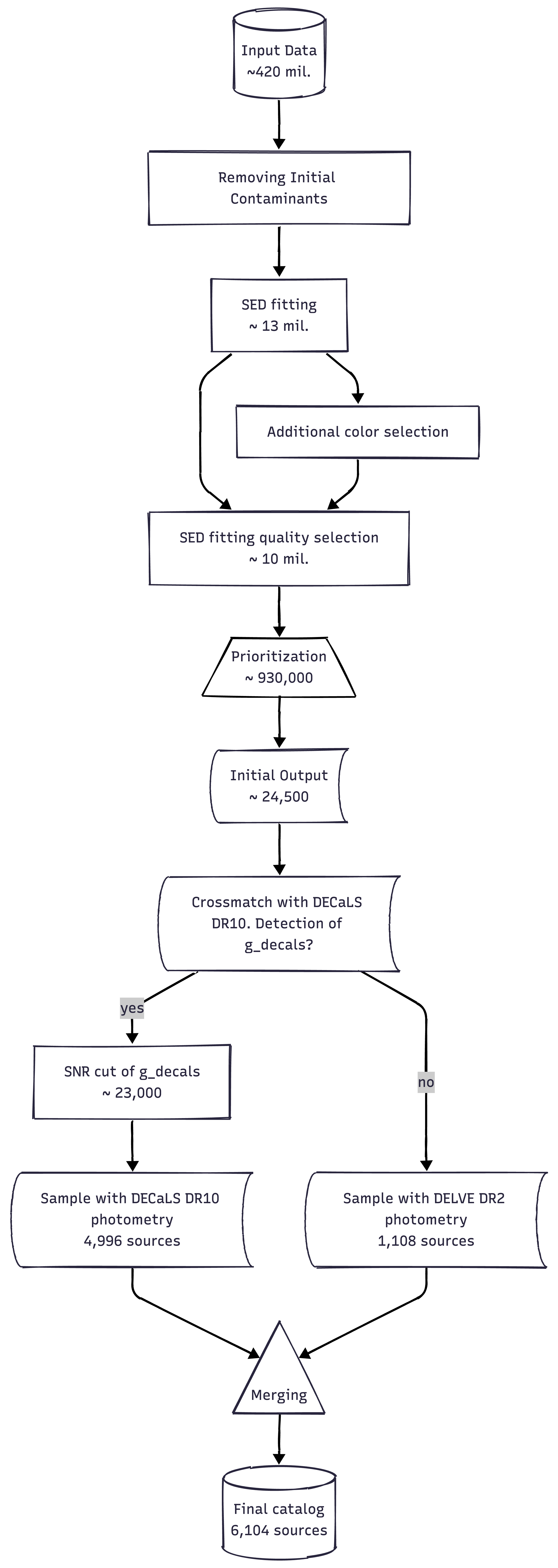}
    \caption{Flowchart of the selection procedure}
    \label{lab:flowchart}
\end{figure}

We impose several initial selection criteria to identify high-redshift quasars at $z > 4.5$, followed by implementing a custom-made SED fitting code and color-color cuts. Finally, we cross-match part of our sample with the DECaLS DR10 catalog to retrieve further information. Our selection procedure is reported schematically in Figure ~\ref{lab:flowchart}, and explained in more detail below.

\subsection{Removing initial contaminants}
\label{sec:removing_contaminants}

Initially, we apply a set of conditions to remove contaminants. To test their effectiveness, we use three high-redshift quasar samples: F23 (a collection of around 500 spectroscopically confirmed quasars at redshifts $ 5.3 < z < 7.6$ from \citealt{Fan_2023}), Y23 (a spectroscopically confirmed list of 556 quasars with redshift values between 4.4 and 6.6, based on DESI data from the study of \citealt{Yang_2023}), and Fl23 (a list of high-redshift quasars from the study of \citealt{Flesch_2023}, totaling 1200 sources with $z>4.5$, out of which we select about 450 spectroscopically confirmed quasars). For comparison with known contaminants, we also use a spectroscopically confirmed sample of brown dwarfs from a collection of studies\footnote{The sample is based on \citep{West_2011,Lodieu_2014, Mace_2014, Marocco_2015, Best_2015}} (totaling 5,700 sources) hereafter referred to as BD. All samples are cross-matched with the surveys adopted in this work. Note that the recovery rate includes sources that passed the selection criteria or those retained due to missing data (NaN values), which pass selection criteria by default.

\begin{figure}[h]
    \centering
    \includegraphics[width=\hsize]{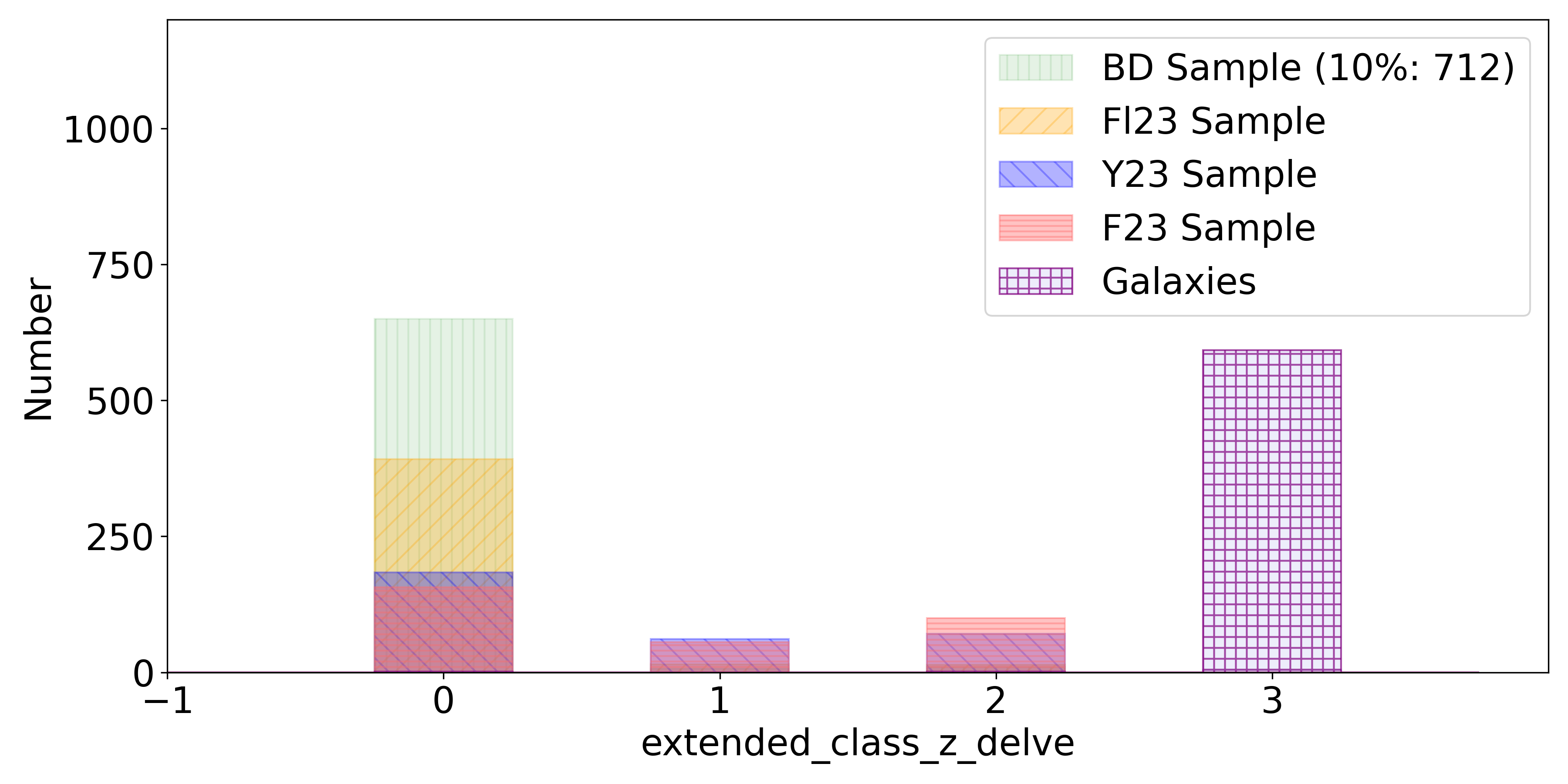}
    \caption{DELVE $extended\_class\_z\_delve$ flag for samples of known high-redshift quasars, randomly selected brown dwarfs from the BD Sample, and galaxies from the literature. The definition and usage of this flag are described in Section~\ref{sec:removing_contaminants}.} 
    \label{fig:data_extz}
\end{figure}

High-redshift quasars are expected to be characterized by a point-like morphology. Since our sample focuses on sources at $z > 4.5$, we used the DELVE survey's $extended\_class\_z\_delve$ flag to distinguish point-like sources from extended ones (e.g., galaxies) based on their $z$-band morphology. Hence, we consider that any high-redshift quasar ($ 4.5 < z < 7$) candidates should be detected in the $z_{delve}$ filter, and choose it as a reference for the morphology of the source. The $extended\_class\_z\_delve$ flag is based on morphological models based on the spread model parameter of {\it Source Extractor} \citep{Bertin_1996} and has the following values: $0$: source has a high likelihood of being a star; $1$: source is probably a star; $2$: source is likely a galaxy; $3$ source has a high likelihood of being a galaxy. Therefore, we excluded all sources with an $extended\_class\_z\_delve$ value of 3, removing obvious extended sources from our dataset. This selection process leads to approximately a 33\% reduction of the initial sample and successfully recovers over 98\% of the high-redshift quasars from all reference samples used. We apply this method to a galaxy sample obtained from the SDSS SkyServer DR10\footnote{\url{https://skyserver.sdss.org/dr10/en/tools/toolshome.aspx}} and find that this criterion eliminates 97\% of them (see Figure ~\ref{fig:data_extz}). 

As high-redshift quasars are not expected to show large proper motion \citep{Heintz_2018}, we eliminate all sources with significant proper motions relative to their measurement uncertainties with the following cut to all sources with matches to Gaia DR3: 

\begin{equation}
\left| \frac{\mu_{\alpha}}{\sigma_{\mu_{\alpha}}} \right| > 2 
\quad \text{or} \quad 
\left| \frac{\mu_{\delta}}{\sigma_{\mu_{\delta}}} \right| > 2,
\end{equation} 

\noindent where $\mu_{\alpha}$ and $\mu_{\delta}$ are the proper motions in RA and Dec, respectively, and $\sigma_{\mu_{\alpha}}$ and $\sigma_{\mu_{\delta}}$ are their corresponding uncertainties.

\begin{figure}[H]
     \centering
    \includegraphics[width=\hsize]{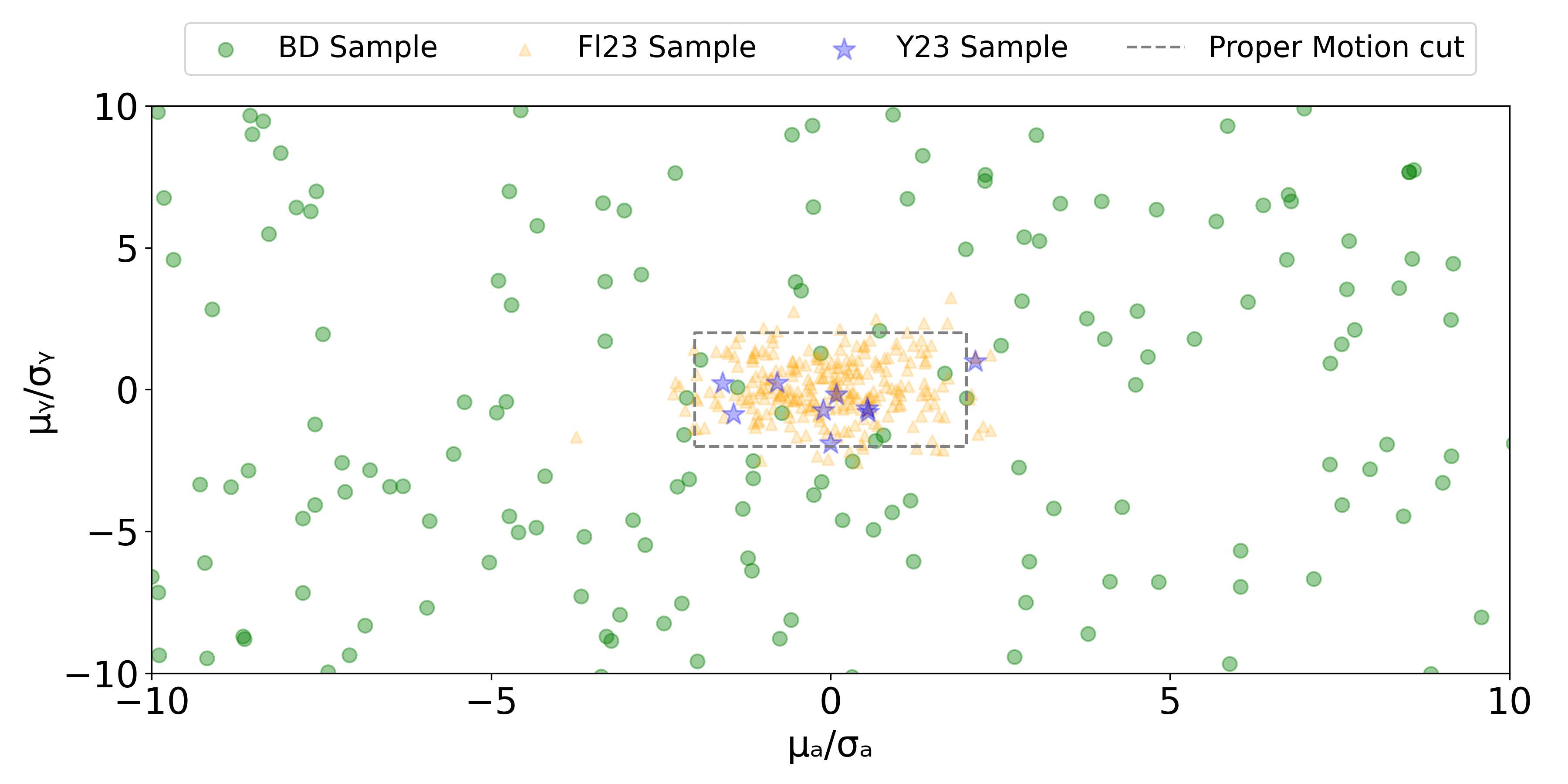}
    \caption{Proper motion significance in declination (\( |\mu_\delta / \sigma_\delta| \)) versus right ascension (\( |\mu_\alpha / \sigma_\alpha| \)) for various source types. BD sample is shown in green, while quasars from Y23 and Fl23 are plotted in blue and orange, respectively. The dashed box outlines the proper motion cut applied in our analysis.}
    \label{fig:data_pm}
\end{figure}

The $2\sigma$ threshold provides an optimal balance between quasar completeness and contamination control, as validated by previous studies of high-redshift quasar selection \citep{Heintz_2018}.
Thus, with this filter, from the sources with matches in Gaia DR3, we retain only 8\% of the BD sample, preserving more than 93\% of the known high-redshift quasars from Y23 and Fl23. No Gaia DR3 counterparts were found for sources in the F23 sample in this analysis (see Figure ~\ref{fig:data_pm}).

\noindent We also exclude all objects with $g_{delve}$ detection, as we aim to select high-redshift quasars at $z>4.5$. Instead, we retain objects with a $g_{delve}$ value of 0 or 99, which may indicate either a lack of detection in the $g$ band or no information. With this cut, we remove 76\% of known brown dwarfs, while retaining over 97\% of known quasars from the F23 and Y23 samples.

Finally, MIR color selection using simple WISE color cuts has been shown to be an effective method for identifying AGN (e.g., \citealt{Stern_2012, DiPompeo_2015, Secrest_2015, Assef_2018}). We apply the following cut on WISE colors in AB.

\begin{equation}
W1 - W2 < 0.6 \quad \text{and} \quad  W1 - W2 > -0.6.
\end{equation}

\begin{figure}[htbp]
     \centering
    \includegraphics[width=\hsize]{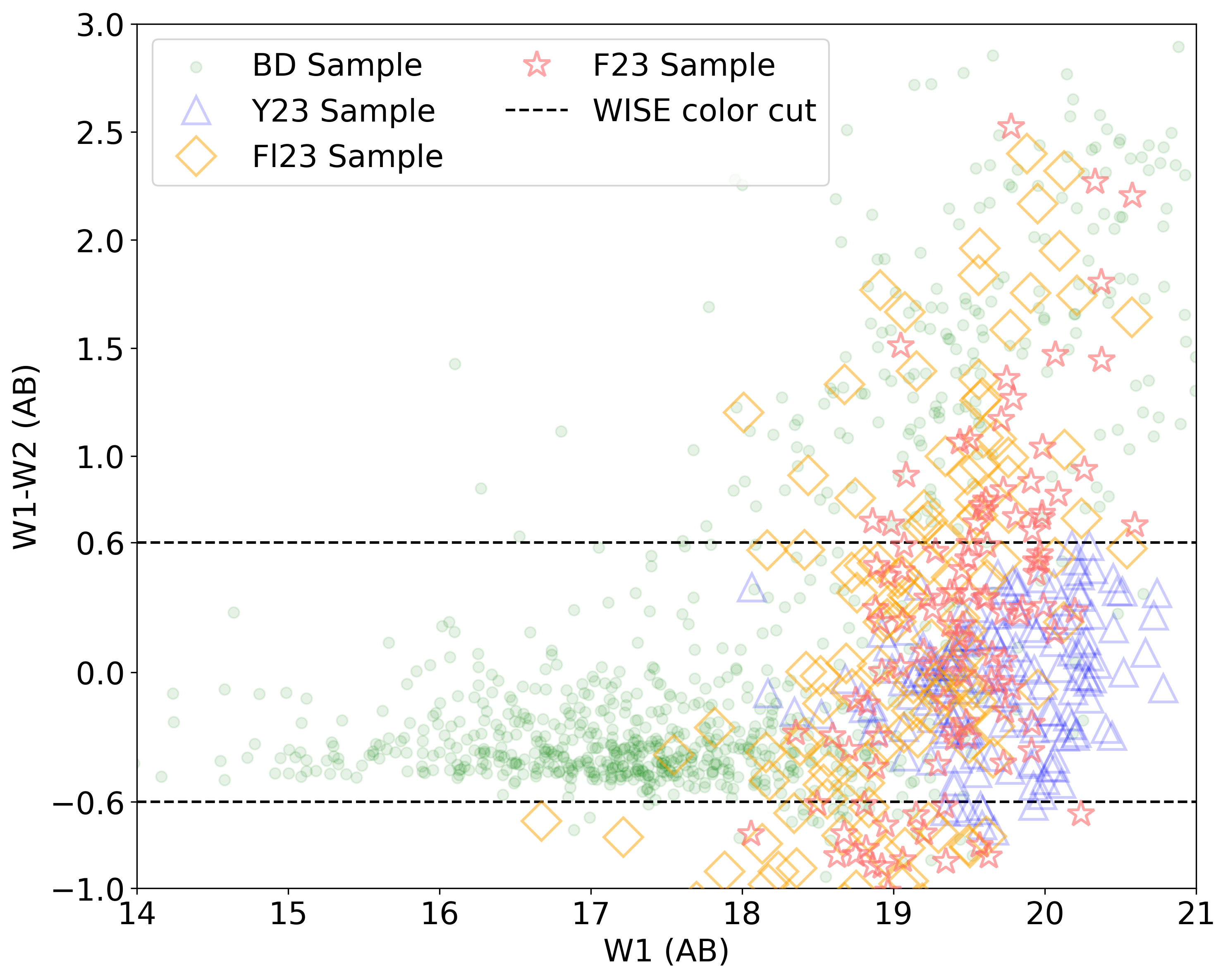}
   \caption{The W1--W2 color criterion is applied to samples of known high-redshift quasars and brown dwarfs. This cut retains more than 76\% of quasars while removing 92\% of brown dwarfs.}
    \label{fig:QSO_BD_wise_colors}
\end{figure}

\noindent This selection keeps over 76\% of known high-redshift quasars matched with ALLWISE from the F23 sample while eliminating approximately 92\% of known brown dwarfs (see Figure~\ref{fig:QSO_BD_wise_colors}). This trade-off prioritizes quasar completeness over brown dwarf rejection, as brown dwarfs significantly outnumber high-redshift quasars. At this stage, we are left with a sample of around 13 million sources, which we will refine further using SED fitting. Table~\ref{tab:Initial_cut}, summarizes the results of the initial cuts performed on the test samples.

\begin{table}[!htbp]
\centering
\scriptsize
\renewcommand{\arraystretch}{1.5}
\caption{Results from the initial cut procedure on known high-redshift quasars and brown dwarfs.}
\label{tab:Initial_cut}
\begin{tabular}{l@{\hspace{0.8em}}c@{\hspace{0.8em}}c@{\hspace{0.8em}}c@{\hspace{0.8em}}c}
\hline
\hline
Conditions & F23 & Y23 & Fl23 & BD \\
 & Recovered & Recovered & Recovered & Remained \\
\hline
\texttt{Extended $z_{delve}$} cut & 321 (99\%) & 318 (98\%) & 445 (99\%) & 3541 (91\%) \\
\texttt{Proper motion} cut & — & 222 (95\%) & 417 (93\%) & 407 (8\%) \\
\texttt{$g_{delve}$} cut & 321 (99\%) & 314 (97\%) & 188 (42\%) & 865 (24\%) \\
\texttt{WISE color} cut & 170 (76\%) & 152 (95\%) & 382 (85\%) & 604 (8\%) \\
\hline
\end{tabular}
\tablefoot{Numbers show recovered sources with recovery percentages in parentheses.}
\end{table}

\subsection{SED fitting}

After applying our initial cuts, we implemented a custom-made SED fitting code \footnote{The SED fitting code is publicly available in github \url{https://github.com/dathevik/high-z-qso-selection.git}} to further refine the sample. As extended sources have been substantially rejected in the previous step, the main contaminants at this stage are brown dwarfs. Indeed, the red flux profile of high-redshift quasars is due to significant absorption of the quasars' ultraviolet emissions by the IGM. On the other hand, brown dwarfs exhibit a gradually rising SED towards 1~\textmu m, due to their surface chemical composition and low temperature \citep[e.g.,][]{Chiu_2007}.  In this step, we use an SED fitting program with various templates. Quasar templates are sourced from \cite{Selsing_2016} and \cite{Temple_2021}, while Brown Dwarf templates are from \cite{Chung_2014}. 

The template described in \cite{Selsing_2016} is based on stacked spectra of luminous quasars observed by the SDSS at a redshift of $z\sim 3$. We shifted the observed template in a redshift range between 4.5 and 6.9, with increments of 0.1, and kept the redshift as a free parameter.  We applied IGM absorption using the method outlined by \cite{Madau_1995}. Additionally, we used a set of synthetic models from the \texttt{qsogen} code\footnote{\url{https://github.com/MJTemple/qsogen/}.} from \cite{Temple_2021}. This model includes three free parameters: emission line strength (weak, normal, and strong), dust reddening characterized by color excess values of $E(B-V)=0, 0.1$, and $0.25$ (where the extinction is attributed to dust within the quasar's host galaxy), and redshift. The redshifts of these synthetic templates range from 4.5 to 7, increasing in steps of 0.05. To account for contamination, four synthetic brown dwarf templates of those used by \cite{Chung_2014} are included, with two free parameters: surface temperature (2500 K, 2000 K, 1500 K, and 1000 K) and gravitational potential (log g = 5.15 and 4.90), based on models from \cite{Allard_2007}. The calculation of the synthetic photometry from all these templates is done with the \texttt{synphot} 1.2.1 python package\footnote{\url{https://synphot.readthedocs.io/en/latest/}.}.

We acknowledge that combining two quasar template sets may, in principle, introduce additional degeneracies; however, we note that $\sim95\%$ of sources in our final catalog are best fit by the \cite{Temple_2021} templates, which include the largest number of free parameters and provide the broadest coverage of quasar properties. The dominant contribution from a single template set, therefore, limits the practical impact of such degeneracies on our SED fitting quality.

We fit the observed SEDs by selecting the BD and quasar templates that minimize the $\chi^2$. We then use the Bayesian Information Criterion $(BIC)$ and $F_{test}$ statistics to discriminate those significantly better fit as quasars. We calculate $\chi^2$ as:

\begin{equation}
\chi^2(z) = \sum_{i=1}^{N_{\text{filters}}} \left[ \frac{F_{\text{obs},i} - b \times F_{\text{temp},i})}{\sigma_i} \right]^2,
\end{equation}

\noindent where \(F_{\text{obs},i} \) and \( F_{\text{temp},i} \) represent the observed and template fluxes, $\sigma_i$ is the flux uncertainty in the filter \textit{i}, and \textit{b} serves as a normalization factor. This normalization factor accounts for the flux normalization of template spectra and ensures optimal matching between observed and template photometry before $\chi^2$ calculation.

\hfill\break
We calculate the $BIC$ as:
\begin{equation}
BIC = \Delta\min\chi^2 + n_{\text{par}}\Delta \times \ln(N_{\text{datapoint}}),
\end{equation}

\noindent where $\Delta\min{\chi^2}$ is the difference between the minimum $\chi^2$ value of the brown dwarf and quasar templates, $n_{\text{par}}\Delta$ is the difference between the numbers of parameters of the brown dwarf and quasar templates and $N_{\rm datapoint}$ is the sum of data points of the source, where a datapoint is the flux in each filter.
 
We use a simplified version of the canonical $F_{test}$ parameter in this study, resulting in the following expression:
 
\begin{equation}
F_{test} = \frac{\left|\min \chi^2\Delta\right| \times k}{n_{\text{par}}\Delta \times \min \chi^2},
\end{equation}

\noindent where ($k$) is the degree of freedom:

\begin{equation}
k = N_{\text{datapoint}} - \left(n_{\text{par \text{model}}} + n_{\text{par}}\Delta\right),
\end{equation}

\noindent and $n_{\text{par \text{model}}}$ is the number of parameters of the applied quasar template.

\begin{figure}[h]
     \centering
    \includegraphics[width=\hsize]{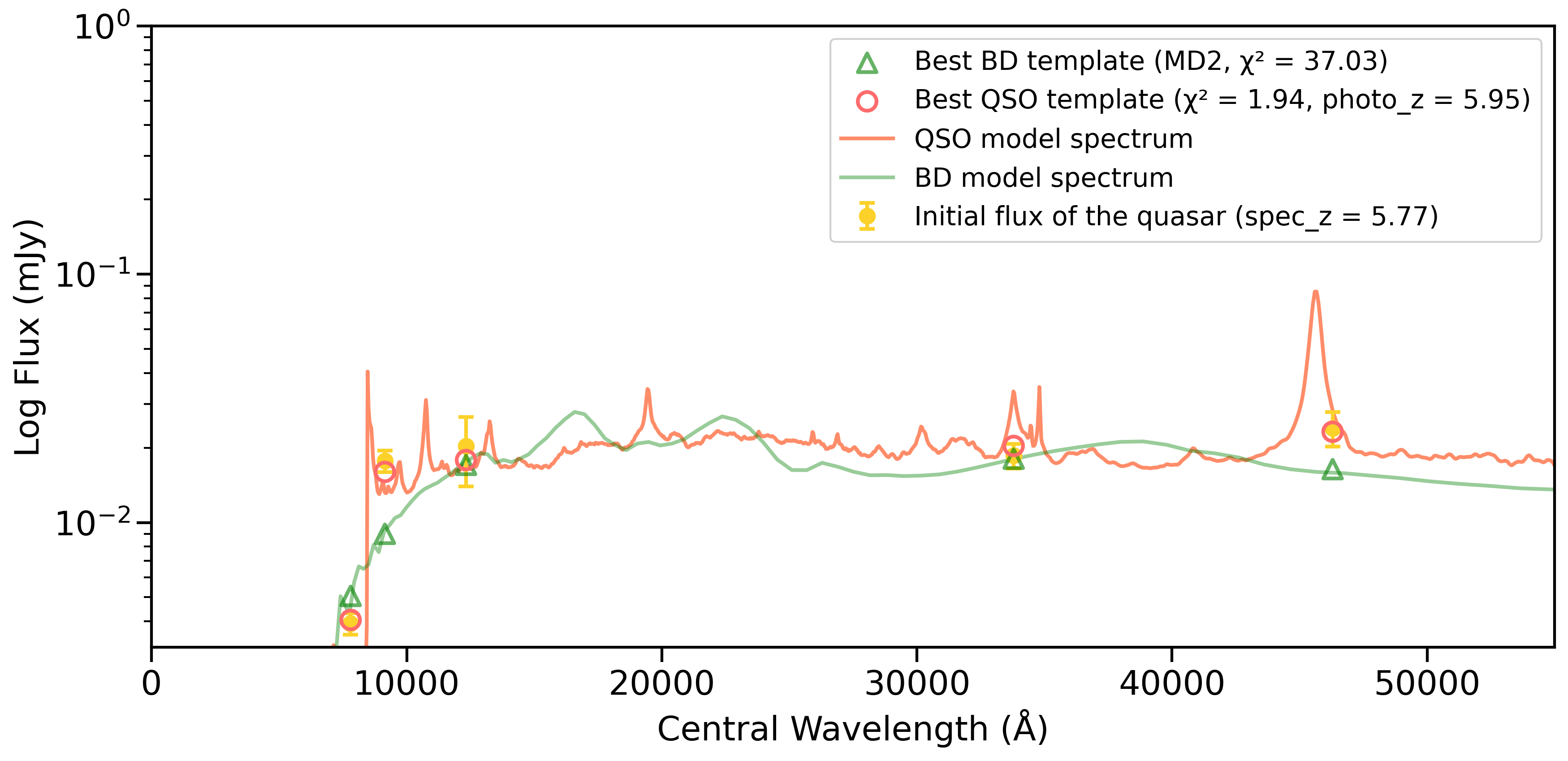}
    \caption{Example of SED fitting for the known quasar $J000009.99-041626.09$ from the F23 sample ($z_{ spec} = 5.77$, $z_{phot} = 5.95$).
 }
    \label{fig:data_sed}
\end{figure}

 \hfill\break
Before performing the SED fitting on our full catalog, we first validate our method using the set of known quasars and brown dwarfs. We assess if the photometric redshift ($z_{phot}$) estimates are suitable for subsequent color selection by comparing them to spectroscopic redshifts ($z_{spec}$) for known quasars, and, most importantly, whether the method successfully distinguishes between quasars and brown dwarf contaminants by examining the ratio of best-fit $\chi^2$ values \citep[e.g.,][]{Ahumada_2020, Fan_2023} (see Figure~\ref{fig:data_sed} for an example). 

\subsection{Additional color selection}
\label{subsec:color-selection}

\begin{figure}[h]
     \centering  \includegraphics[width=\hsize]{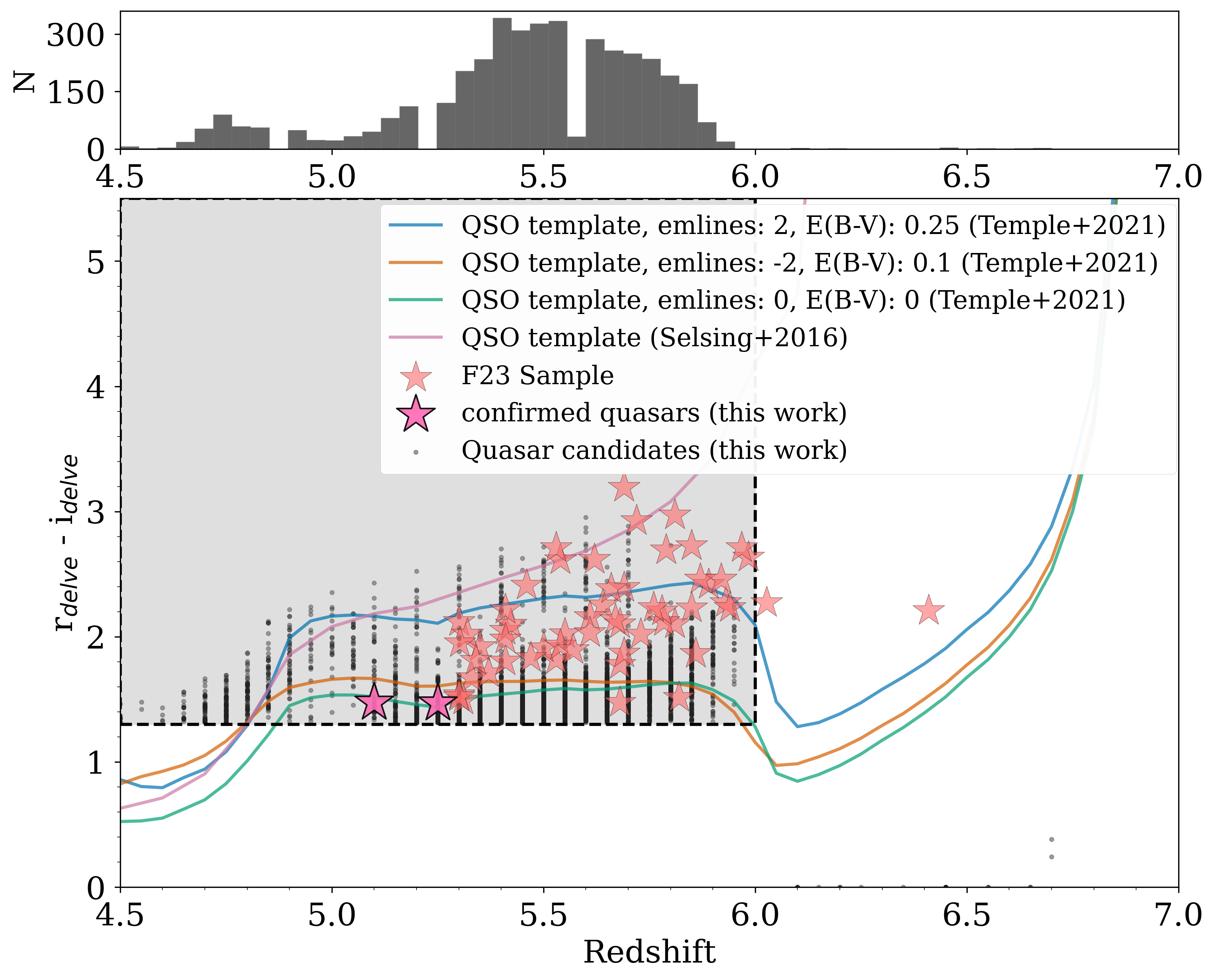}
    \caption{The relation between $r_{delve} - i_{delve}$ color and the redshift for high-redshift quasars from the F23 sample, using DELVE photometry, and our  quasar candidates (this work), marked as small gray dots. The histogram at the top shows  the distribution of SED-based photometric redshifts of our quasar candidates. The shaded gray area indicates the selection region defined by the criterion  $r_{delve} - i_{delve} > 1.3$, which is applied to sources at $z_{phot} < 6$. All sources at $z_{phot} > 6$ are retained regardless of their color. The 
confirmed quasars from this work are shown as large pink stars.}
    \label{fig:data_color}
\end{figure}

At this stage, following the SED fitting step, we have approximately 13 million candidates. We use the SED-calculated $z_{phot}$ to perform additional color-based selection this time with $r_{delve}-i_{delve}$ colors from DELVE. We use the color distribution of known high-redshift quasars from F23 sample to determine the optimal threshold for the $r_{delve}-i_{delve}$ bands and find that sources at $z_{spec} > 4.5$ systematically exhibit $r_{delve} - i_{delve} > 1.3$ (see Figure~\ref{fig:data_color}). Therefore, for objects with $z_{phot} < 6$, we imposed a color cut of $r_{delve} - i_{delve} > 1.3$. Because of the small number of sources (178) at $z_{phot} > 6$ in our sample, the color cut is performed only on the sources with $z_{phot} < 6$, while all the sources with $z_{phot} > 6$ are retained without applying color cuts. This cut removed approximately 3 million sources, retaining 10 million for further cuts.

\subsection{SED fitting quality selection}
\label{subsec:quality-selection}

\begin{figure}[h]
     \centering    \includegraphics[width=\hsize]{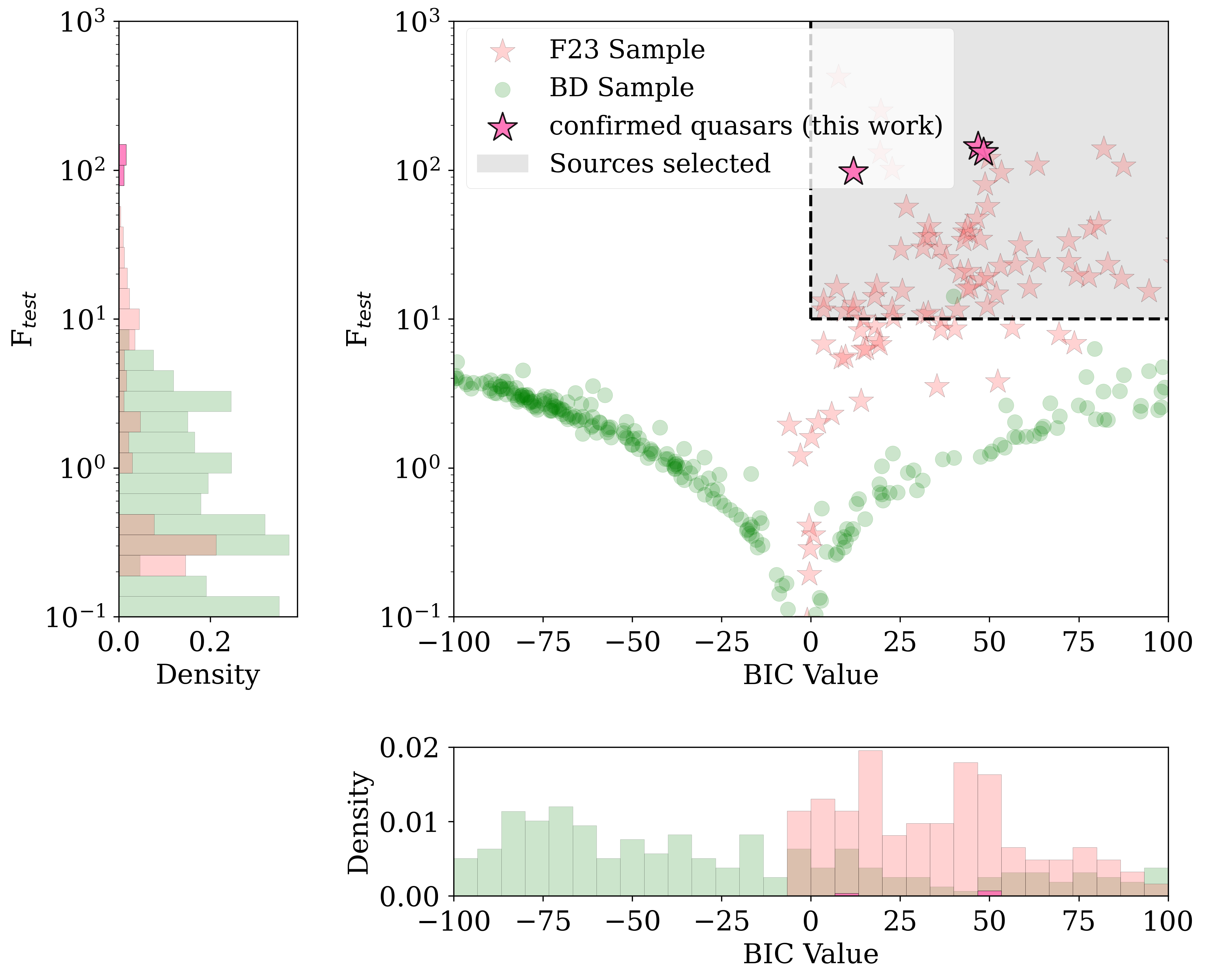}
    \caption{The distribution of $F_{test}$ and $BIC$ values of the high-redshift quasars from the F23 sample, the BD sample and the spectroscopically confirmed high-redshift quasars from our sample. The shaded gray area indicates the selection region defined by our criteria ($BIC > 0$ and $F_{test} > 10$). The subplots indicate the density histogram of the corresponding axis.}
    \label{fig:stat_results}
\end{figure}

At this stage, we consider the $\chi^2$, $BIC$, and $F_{\mathrm{test}}$
values obtained from our SED fitting from the F23 and BD samples (see Figure~\ref{fig:stat_results}). To test the effectiveness of each criterion, we tested them separately on known brown dwarfs: $BIC > 0$ alone removes around 57\%, while $F_{\mathrm{test}} > 10$ alone removes approximately 99\% of the BD sample. In our selection pipeline, we apply both criteria simultaneously to maximize contaminant rejection. Additionally, to ensure we select candidates with reliable and genuinely multiwavelength photometry, we require sources to have detections in 7 or more photometric bands, which typically guarantees coverage across all 3 surveys or across 2 surveys with substantial photometric information. For sources with 7--8 bands, missing data are most commonly due to the gaps in VHS/VIKING fields in $H$ or $Y$ bands, and more rarely in the DELVE $r$ band coverage.

To summarize, we require: 

\begin{itemize}
  \item $BIC > 0$ AND $F_{test}
 > 10$
  \item $N_{\rm datapoint} > 6$
\end{itemize}

\subsection{Prioritization}
\label{subsec:prioritization}

At this stage, we have narrowed our sample to approximately 930000 candidates. After performing morphology, color, SED fitting, and related cuts, we prioritize sources for follow-up spectroscopy based on their brightness, their respective $\chi^2_{\text{QSO,min}}$, and $R_{\chi²}$ values defined as:

\begin{equation}
R_{\chi^2} = \frac{\chi^2_{\text{QSO,min}}}{\chi^2_{\text{BD,min}}}
\label{eq:R_chi},
\end{equation}

\noindent where $R_{\chi^2} < 1$ indicates a better fit with quasar templates than brown dwarf templates. This criterion alone removes 56\% of known brown dwarfs while preserving more than 94\% of known quasars.

We prioritize our candidates according to the following criteria
\begin{itemize}
\item Priority 2: $z_{\text{delve}} \leq 22$ 
\item Priority 1: $z_{\text{delve}} \leq 22$ AND $\chi^2_{\text{QSO,min}} \leq 1.2$ AND $R_{\chi^2} \leq 0.3$ 
\end{itemize}

\noindent Additionally, sources within $5^\circ$ of the Large Magellanic Cloud and $2.5^\circ$ of the Small Magellanic Cloud are excluded to avoid crowded regions, and the sample is restricted to declinations between $-80^\circ$ and $5^\circ$ to focus on observable targets (Bauer et al. in prep).

After this cut, we narrowed our initial output catalog to approximately 24,500 sources with priority 1 and 2. 

\subsection{Sample with DECaLS DR10 photometry} 
\label{sec:decals_sample}
We cross-matched the remaining $\sim$24,500 sources with the DECaLS DR10 catalog via the NOIRLab Datalab interface\footnote{https://datalab.noirlab.edu}, to obtain further photometric information. Of these, over 23000 have counterparts in $g_{decals}$. We applied a $3\sigma$ detection threshold criterion, which reduced our candidate list to 4996 objects. This is motivated by the fact that quasars at $z > 4.5$ are not expected to be detected in the $g$-band due to Lyman-break absorption, and the deeper $g$-band photometry of DECaLS DR10 ($g_{decals} = 24.7$) compared to DELVE DR2 ($g_{delve} = 24.3$) provides a strict criterion for excluding low-redshift contaminants with $g-band$ detections. As a result, we adopted DECaLS photometry for sources with $SNR\_g_{decals} < 3$ and kept DELVE DR2 magnitudes for 1108 candidates without matches in DECaLS DR10, due to the incomplete sky coverage of DECaLS DR10. This approach creates two samples: one containing $griz$ bands from DELVE DR2  and the other from DECaLS DR10 selected based on the SNR of $g_{decals}$ (see Figure~\ref{lab:flowchart} for the complete selection layout). Our final catalog of high-redshift quasar candidates, which will be observed within ChANGES, includes 6104 sources in total of which 58\% are priority 1 sources. 

\section{Results}
\label{sec:results}
\begin{figure}[h]
     \centering
    \includegraphics[width=\hsize]{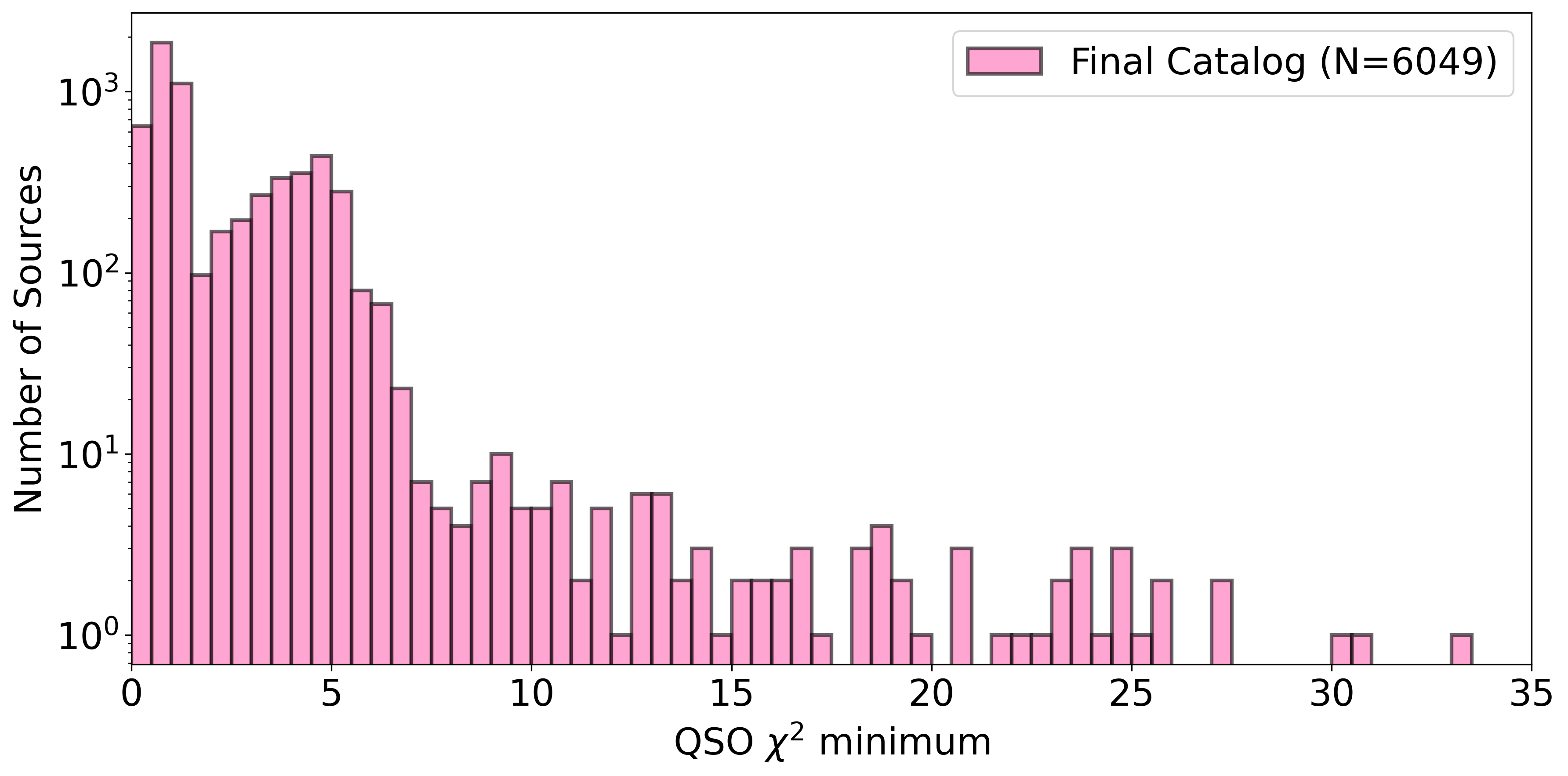}
    \caption{The $\chi^2_{\text{QSO,min}}$ distribution of the final catalog of 6104 high-redshift quasar candidates. The distribution shows the first sharp peak near $\chi^2_{\text{QSO,min}}\sim 1$ due to the performed prioritization cuts.}
    \label{fig:results_chi2}
\end{figure}

Our main aim is to identify a sample of high-redshift quasar candidates for the ChANGES survey. In Figure~\ref{fig:results_chi2}, we report the distribution of the minimum \( \chi^2 \) values obtained from SED fitting with quasar templates for our final catalog, which includes sources classified as both priority 1 and priority 2 candidates (see Section~\ref{subsec:prioritization}). Given that all sources in the sample are required to have photometric information in 7 or more filters, and that the fitting procedure involved either 3 or 1 free parameters, depending on the best fitting template, the expected acceptable range for total $\chi^2$ values is approximately between 4 and 14. This corresponds to reduced $\chi^2$ values (i.e., $\chi^2_\nu = \chi^2/\nu$, where $\nu$ is the number of degrees of freedom) ranging from $\sim$1 to $\sim$5, where lower values indicate statistically better fits.

To test the completeness of our sample of high-redshift quasar candidates, we chose all the available sources in declinations between $-30^\circ$ and $0^\circ$ to get the highest number of known high-redshift quasars from samples of F23. As a control variable, we kept the sources with $z_{\text{delve}} < 22$. The completeness and reliability are estimated using the formulas:

\begin{equation}
C = \frac{\text{Number of selected quasars}}{\text{Total number of confirmed quasars}}
\label{eq:completeness},
\end{equation}

\noindent where completeness (C) measures how well our selection method recovers known quasars from the total population of confirmed quasars from the F23 sample in our selected region; and

\begin{equation}
R = \frac{\text{Number of selected quasars}}{\text{Total number of selected objects}},
\label{eq:purity}
\end{equation}

\noindent reliability (R) quantifies the fraction of our selected candidates that are quasars, providing a measure of contamination in our sample. 

\begin{figure}[!htbp]
    \centering
    \includegraphics[width=\hsize, keepaspectratio]{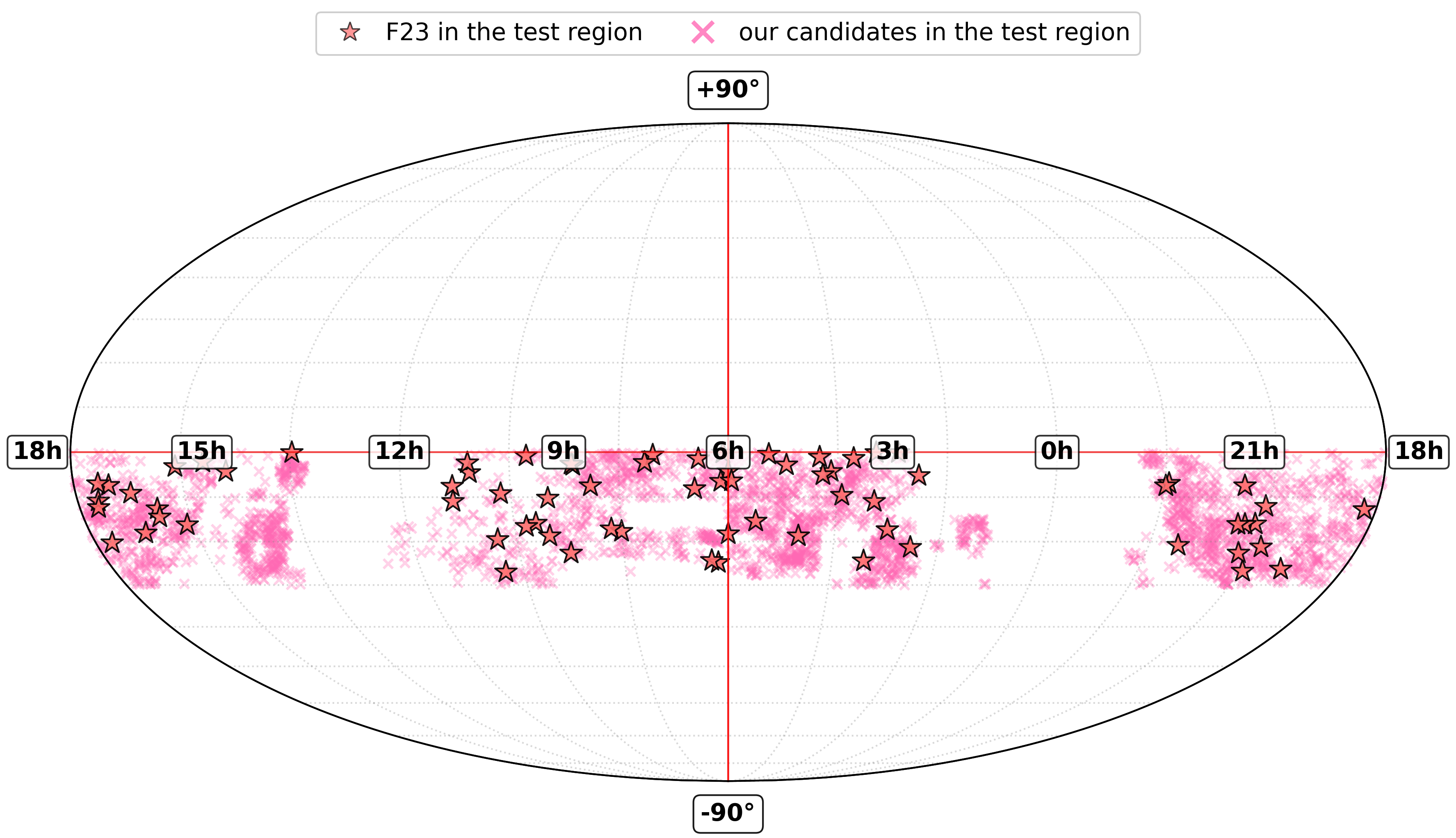}
    \vspace{0.5cm} 
    \includegraphics[width=\hsize, height=13cm, keepaspectratio]{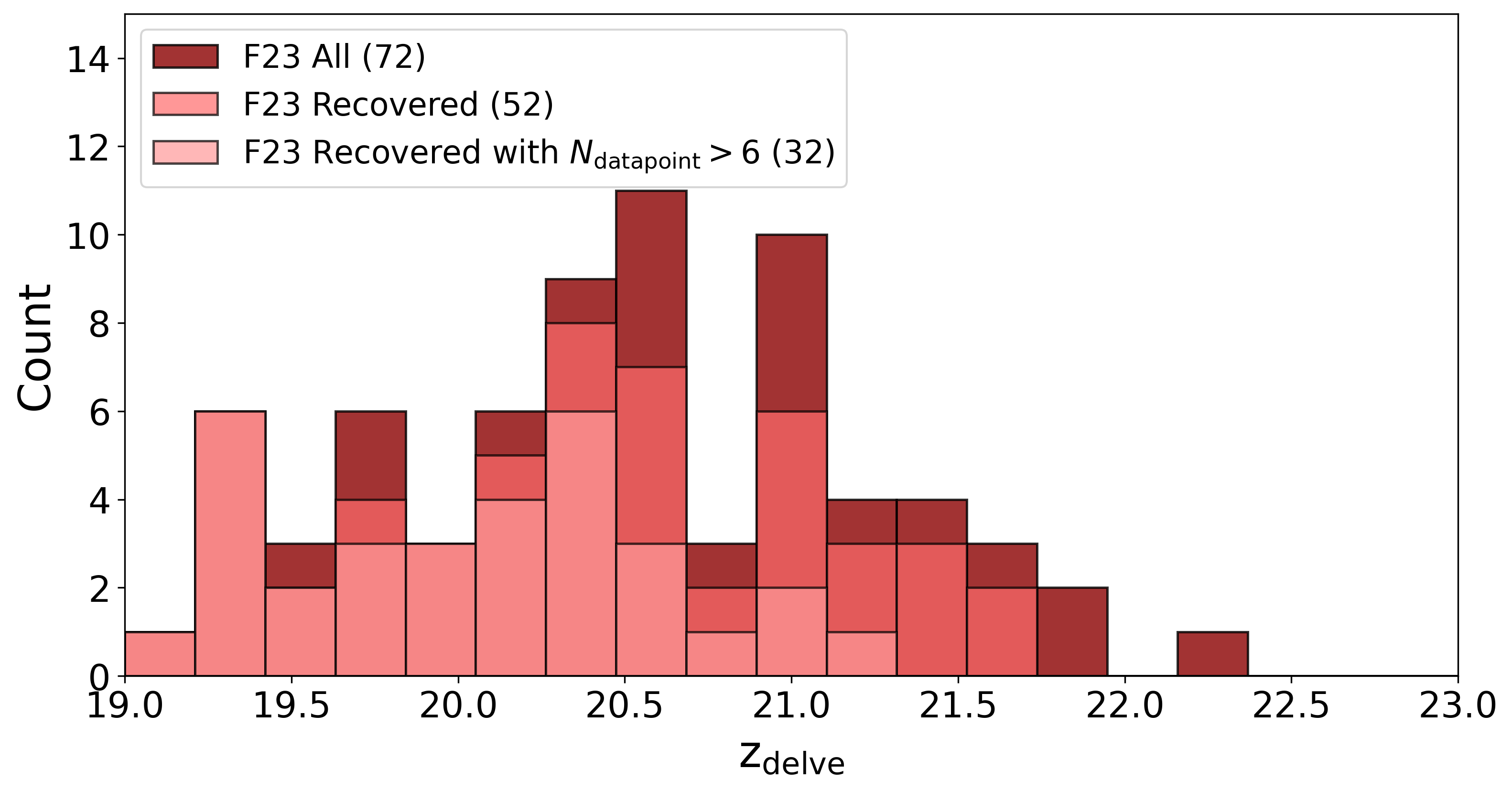}
 \caption{Distribution of sources from the F23 sample and our final catalog within the declination range $-30^{\circ} < \text{Dec} < 0^{\circ}$. 
Top panel: Sky distribution of the selected region for our candidates of high-redshift quasars with high-redshift quasars from the F23 sample. Bottom panel: The histogram of $z_{\text{delve}}$ for the F23 sample in the selected region and with respect to the ones recovered after the selection procedures with and without limiting the $N_{\text{datapoint}}$.} 
   \label{fig:completeness}
\end{figure}

With this approach, we reach 44\% completeness and 1.5\% reliability (see Figure ~\ref{fig:completeness}). Here, reliability is computed as the fraction of our selected candidates that are spectroscopically confirmed quasars from the F23 sample; the remaining candidates have not been spectroscopically confirmed and should not be assumed to be contaminants. The pink points in the top panel of Figure ~\ref{fig:completeness} represent our unconfirmed quasar candidates. Therefore, given the incompleteness of southern sky spectroscopic samples, the measured reliability should be interpreted as a conservative lower limit rather than a definitive estimation of our selection's true performance. A full assessment of completeness and reliability would require modeling the selection function to quantify the impact of each selection criterion, which is deferred to future work after dedicated spectroscopic follow-up.

\begin{figure}[!htbp]
     \centering
\includegraphics[width=\hsize,height=22 cm, keepaspectratio]{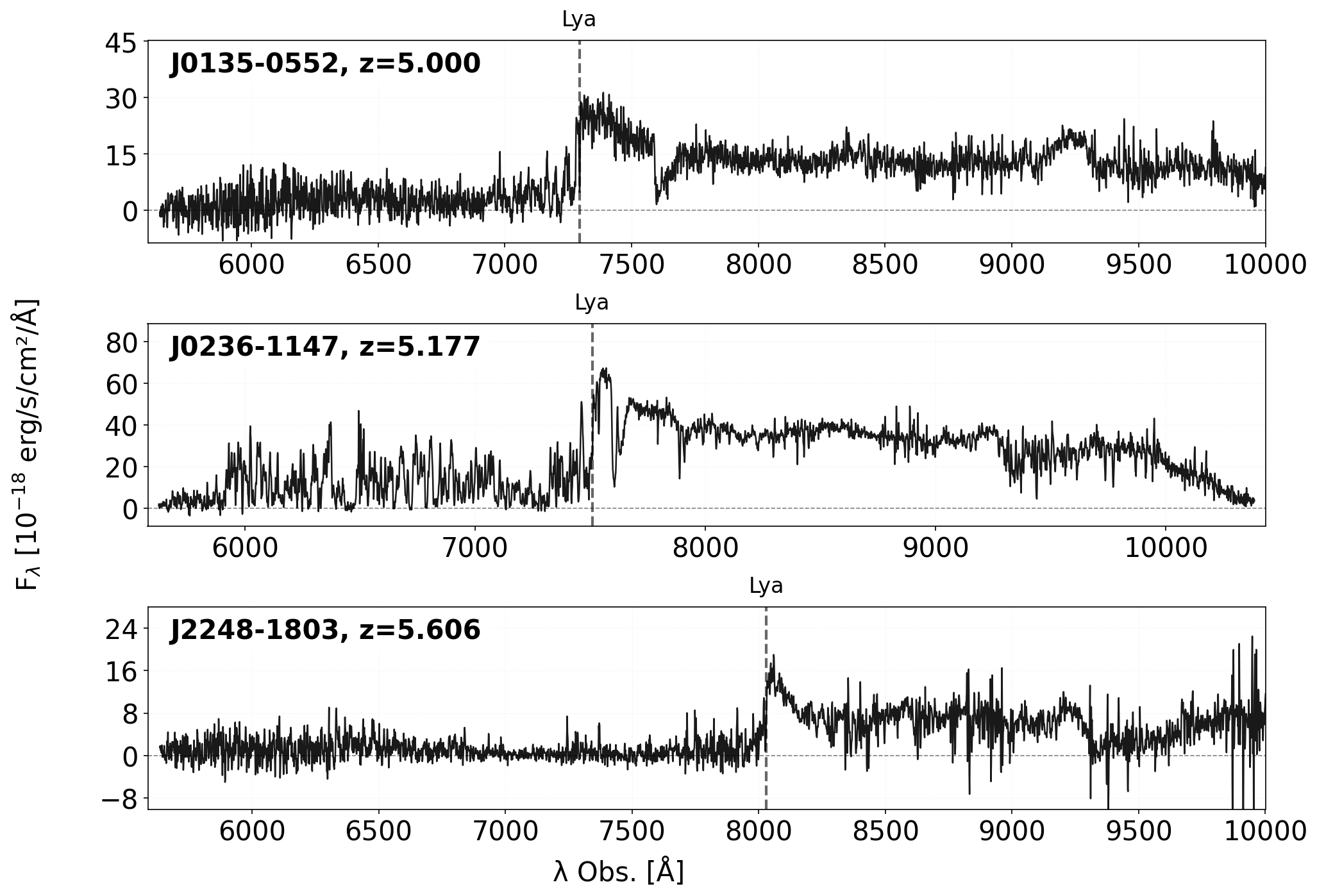}
    \caption{Palomar/NGPS spectrum of the 3 confirmed high-redshift quasars, observed on UT 2025 August-November. The spectra shows the   Lyman-$\alpha$ break (indicated by the vertical dashed line), confirming the quasar nature for those candidates selected through our SED fitting methodology.}
    \label{fig:palomar_spectrum}
\end{figure}

As part of our validation efforts, six high-redshift quasar candidates from our final catalog were observed spectroscopically. Initial observations of three candidates were conducted with the New Technology Telescope (NTT) at La Silla Observatory during ESO program ID 115.2816.001 between UT 2025 July 27 and August 2. The observations were conducted using the ESO Faint Object Spectrograph and Camera \citep[v.2; EFOSC2][]{Buzzoni_1984} with the Grism 5 (5200--9350~\AA), which provided spectral coverage suitable for confirming high-redshift quasars by detecting the Lyman break. The observing run faced challenges due to poor weather conditions, with variable seeing ranging from $1.1-1.8\arcsec$ and intermittent cloud cover. One source (\texttt{Name}:/$J2248-1803$, $z_{delve} = 19.95$) showed a promising Lyman-break at $z \sim 5.6$, representing a potential high-redshift quasar. The other two targets (\texttt{Name}:/$J2223-1042$ and $J2228-1650$) were identified as contaminants: one is a low-redshift ($z\sim3.2$) source, and the other showed inconclusive results. The data for all sources were reduced with standard routines \citep[e.g.,][]{Belladitta_2025} using the \texttt{PypeIt} pipeline\footnote{\url{https://pypeit.readthedocs.io/en/stable/}} \citep{Prochaska_2020}.
\noindent Given the promising but very low signal-to-noise ratio NTT/EFOSC2 spectrum of J2248-1803 and the poor observing conditions (seeing up to $1.8\arcsec$), further follow-up observations were conducted with the Palomar Observatory using the Next Generation Palomar Spectrograph \citep[NGPS;][]{Jiang_2018} on UT 2025 August 23. The spectrum of the source clearly shows the characteristic Lyman-break at $\sim 8000$~\AA, confirming the quasar at a spectroscopic redshift of $z_{spec} = 5.606$, consistent with the photometric redshift estimate from our SED fitting analysis. Following the confirmation of J2248-1803, three additional candidates were observed with Palomar/NGPS. The Palomar/NGPS spectra were reduced using standard IRAF routines. Of these three targets, two were successfully confirmed as high-redshift quasars (see Figure~\ref{fig:palomar_spectrum}). The third target (Object ID 10735100050971)is likely an M-type star (see Table~\ref{tab:spectroscopic_followup}). Overall, our spectroscopic campaign achieved a detection rate of 50\% (3 confirmed quasars out of 6 observed candidates). However, we note that this rate is not representative of the full catalog, as these candidates were specifically selected based on their brightness and visual inspection. The true reliability of our selection will be determined once spectroscopic information is obtained for the entire catalog.
\noindent It is important to note that all these sources were selected from our catalog of high-redshift quasar candidates based on their $z$-band magnitudes ($z_{mag,decals} < 20.5$). We also visually inspected their images in the Legacy Survey Sky Viewer and PanSTARRS-1 Image Access to prioritize bright and clear point-like sources.

\begin{table*}[h]
\centering
\scriptsize
\renewcommand{\arraystretch}{1.3}
\caption{Information on the six high-redshift quasar candidates spectroscopically followed up with NTT/EFOSC2 and Palomar/NGPS.}
\label{tab:spectroscopic_followup}
\begin{tabular}{l@{\hspace{0.5em}}c@{\hspace{0.5em}}c@{\hspace{0.5em}}c@{\hspace{0.5em}}c@{\hspace{0.5em}}c@{\hspace{0.5em}}c@{\hspace{0.5em}}c@{\hspace{0.5em}}c@{\hspace{0.5em}}c@{\hspace{0.5em}}c@{\hspace{0.5em}}c@{\hspace{0.5em}}c@{\hspace{0.5em}}l}
\hline
\hline
Name & RA & Dec & $g_{decals}$ & $r_{decals}$ & $i_{decals}$ & $z_{decals}$ & $\chi^2_{QSO}$ & $R_{\chi^2}$ & $z_{phot}$ & $z_{spec}$ & Exp. Time & Seeing & Telescope/Instrument \\
& (deg) & (deg) & (AB mag) & (AB mag) & (AB mag) & (AB mag) &  &  &  &  & (s) & (arcsec) &  \\
\hline
$J2228-1650$ & 337.0605 & -16.8485 & — & — & $19.97 \pm 0.010$ & $19.71 \pm 0.013$ & 0.589 & 0.013 & 5.350 & — & 1800 & 1.2 & NTT/EFOSC2 \\
$J2248-1803$ & 342.1769 & -18.0430 & $26.99 \pm 2.38$ & — & $20.79 \pm 0.015$ & $19.96 \pm 0.018$ & 2.048 & 0.012 & 5.650 & 5.606 & 1800 & 1.2 & Palomar/NGPS \\
$J2109-2639$ & 317.4431 & -26.6500 & — & — & $20.28 \pm 0.016$ & $19.85 \pm 0.020$ & 1.471 & 0.030 & 5.150 & — & 1800 & 1.2 & NTT/EFOSC2 \\
$J0135-0552$ & 23.7621 & 5.8731 & $25.90 \pm 0.27$ & $21.70 \pm 0.05$ & $20.17 \pm 0.02$ & $20.00 \pm 0.02$ & 1.354 & 0.027 & 5.100 & 5.000 & 1200 & 1.3 & Palomar/NGPS \\
$J0237-1148$ & 39.2023 & 11.7926 & $26.97 \pm 0.27$ & $20.22 \pm 0.03$ & $18.75 \pm 0.01$ & $18.45 \pm 0.01$ & 1.527 & 0.029 & 5.250 & 5.177 & 1200 & 1.0 & Palomar/NGPS \\
$J2223-1042$ & 335.7022 & 10.7067 & $25.10 \pm 0.15$ & $23.23 \pm 0.10$ & — & $20.41 \pm 0.03$ & 2.138 & 0.029 & 6.100 & — & 1200 & 1.0 & Palomar/NGPS \\
\hline
\end{tabular}
\end{table*}

\section{Summary}
\label{sec:summary}
\begin{figure}[h]
     \centering
    \includegraphics[width=\hsize]{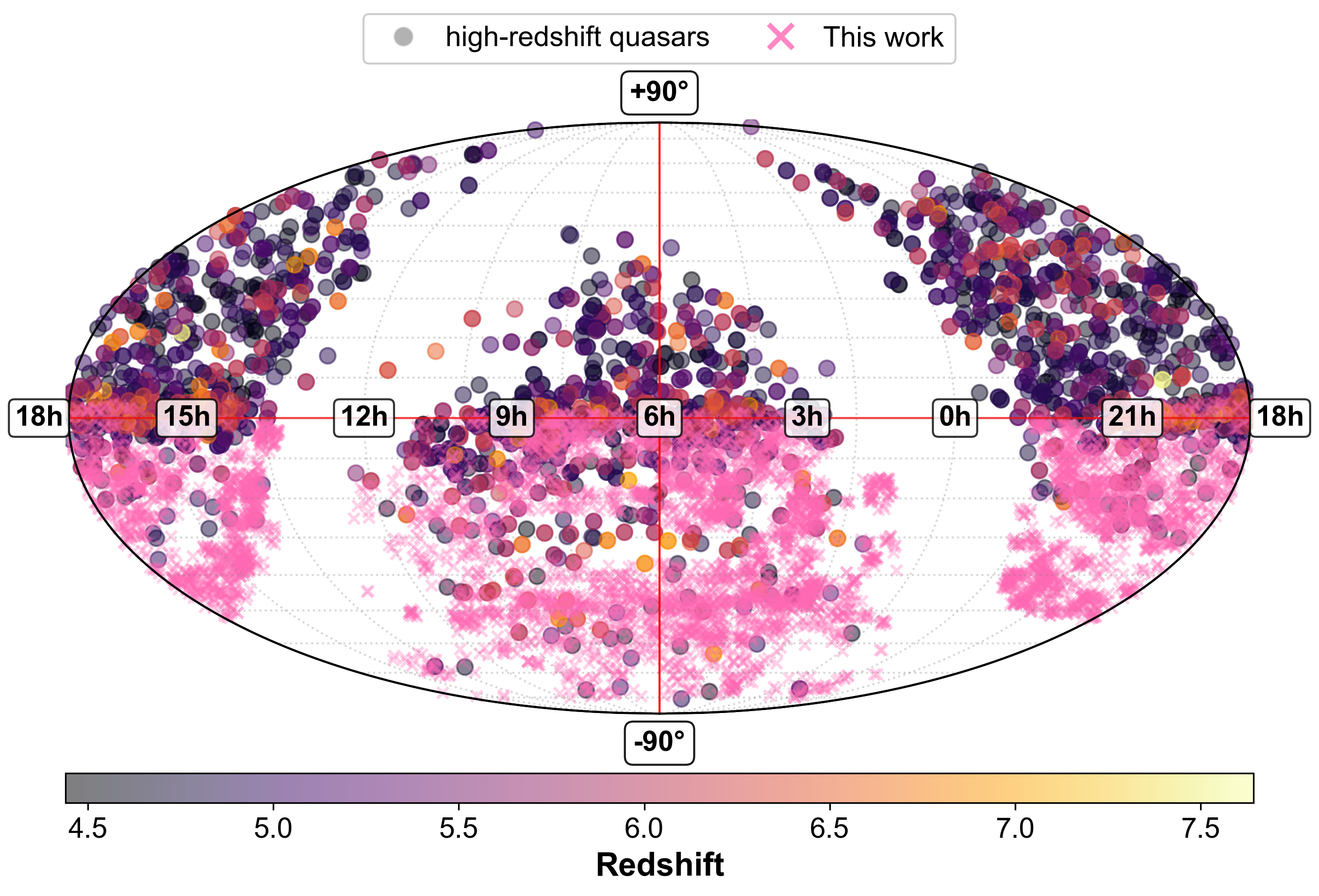}
    \caption{Spatial distribution of high-redshift quasar candidates in the southern hemisphere. Pink crosses show our candidates, while darker points are spectroscopically confirmed high-redshift quasars from the literature (F23, Y23, Fl23 samples). The deficit around RA $\sim 120^\circ$--$240^\circ$ results from increased Galactic extinction and our $E(B-V) < 0.3$ selection criterion. This distribution aligns well with the ChANGES survey strategy, which targets $\sim 18,000~\mathrm{deg}^2$ across declinations between $-80^\circ$ and $+5^\circ$ to maximize overlap with future LSST observations and match 4MOST's optimal observing range.}
    \label{fig:data_visualization}
\end{figure}

In this work, we developed multistep selection procedures to identify high-redshift quasar candidates for spectroscopic follow-up as part of the 4MOST ChANGES survey. Starting from a cross-matched catalog of $\sim420$ million sources across the southern hemisphere, we applied a series of morphological, color-based, and statistical criteria, followed by custom-built SED fitting using empirical and theoretical templates for both quasars and brown dwarfs. Our initial filtering reduces the candidate list to approximately 25,000 sources, from which a final catalog of 6104 candidates is created (see Figure~\ref{fig:data_visualization}).

\noindent As shown in Table~\ref{tab:QSO_BD_cut}, the sequential application of selection criteria significantly narrows the candidate pool. For example, the color-color selection ($r_{delve} - i_{delve} > 1.3$ criterion applied in Section~\ref{subsec:color-selection}), retains a large fraction of known quasars (e.g., 100\% for the F23 sample). However, subsequent steps, including the data quality requirements (described in the Section~\ref{subsec:quality-selection}) and the $E(B-V) < 0.3$ dust extinction cut, reduce the retained sample to below 40\% across the known high-redshift quasar samples. The prioritization criteria (Section~\ref{subsec:prioritization}) further refine the $\sim$930000 remaining candidates by ranking them on brightness and SED fit quality, recovering 7--8\% of known quasars in the higher-confidence tier. Finally, the $N_{\rm datapoint} > 6$ requirement, which is also a part of the quality selection in Section~\ref{subsec:quality-selection}, ensures reliable multiwavelength coverage, reducing the recovered fractions to 26--37\%. We note, that the $N_{\rm datapoint}$ and $z_{\rm delve}$ brightness cuts serve as data quality filters rather than astrophysical discriminators, and are therefore listed below the dashed line in Table~\ref{tab:QSO_BD_cut}.  

\begin{table}[!htbp]
\centering
\scriptsize
\renewcommand{\arraystretch}{1.2}
\caption{Results of the selection process to identify high-redshift quasars and brown dwarfs.}
\label{tab:QSO_BD_cut}
\begin{tabular}{l@{\hspace{0.8em}}c@{\hspace{0.8em}}c@{\hspace{0.8em}}c@{\hspace{0.8em}}c}
\hline
\hline
Conditions & F23 & Y23 & Fl23 & BD \\
 & Recovered & Recovered & Recovered & Remained \\
\hline
Color-color selection & 100\% & 96\% & 32\% & 98\% \\
$BIC>0$ and $F_{test}>10$ & 70\% & 39\% & 65\% & 0.004\%\\
$E(B-V)<0.3$ & 70\% & 38\% & 70\% & 0.004\%\\
Priority 1 & 7\% & 8\% & 9\% & — \\
\hdashline
Priority 2 & 65\% & 38\% & 94\% & 0.001\% \\
$N_{datapoint} > 6$ & 37\% & 28\% & 26\% & 0.0008\% \\
\hline
\end{tabular}
\tablefoot{The percentages indicate the fraction of sources retained after each selection criterion with respect to the total number of detected sources. See Section~\ref{sec:method} for more details and references for each sample. The selection criteria above the dashed line represent the core scientific filters based on statistical, morphological, and color criteria, achieving recovery rates of 70\% for high-redshift quasars in the F23 sample (with 7\% being higher priority sources) and 38\% for the Y23 sample (with 8\% being higher likelihood sources) and 18\% for the Fl23 sample (with 9\% being higher likelihood sources). The criteria below the dashed line ($z_{\text{delve}}$ cut for Priority 2 and $N_{\text{datapoint}}$) are applied to ensure high data quality. They are not included in the final recovery rate calculations as they filter for source trustworthiness rather than astrophysical properties.}
\end{table}

\noindent
The performance of our selection methodology can be assessed through three key metrics: recovery rate, completeness, and reliability. Our recovery rates from known high-redshift quasar samples (see Table~\ref{tab:QSO_BD_cut}) demonstrate that 7-8\% of confirmed quasars are retained after our complete selection process. Our completeness assessment, focusing on sources with optimal data quality in the declination range $-30^\circ < Dec < 0^\circ$, yields 44\% recovery of known quasars, indicating that our selection criteria identify a substantial fraction of the existing high-redshift quasar population.The reliability of our sample, defined as the fraction of selected candidates that are 
quasars without imposing the $N_{datapoint} > 6$ requirement, reaches 1.5\% 
in our test region. Assuming a similar population in our target field, and considering that our input catalog contains over 6,000 candidates, we expect to identify about 7\% of high-redshift quasars. However, this estimate should be approached with caution, as the actual number will depend on the properties of the sources within the observed field and how representative the comparison catalogs are.  

\noindent
The catalog presented in this work will be observed with 4MOST at the beginning of 2026. It consists of 44 parameters (see Table~\ref{tab:outputcatalog_columns}), including magnitudes across various photometric bands, statistical results from the SED fitting, and $E(B-V)$ values from SFD maps. 

\section*{Data availability}
The full catalog of high-redshift quasar candidates is only available 
in electronic form at the CDS via anonymous ftp to 
\texttt{cdsarc.u-strasbg.fr} (130.79.128.5) or via 
\url{http://cdsweb.u-strasbg.fr/cgi-bin/qcat?J/A+A/}.

\begin{table*}[!htbp]
\centering
\large
\renewcommand{\arraystretch}{1.3}
\caption{Column-by-column description of the high-redshift quasar candidates catalog.}
\label{tab:outputcatalog_columns}
\begin{tabular}{l@{\hspace{1.2em}}p{12cm}}
\hline
\hline
Column & Description \\
\hline
\texttt{object\_id} & ID number of the source taken from DELVE DR2 \\
\texttt{ra}, \texttt{dec} & Right ascension and declination (degrees) taken from DELVE DR2 \\
\texttt{mag\_g\_delve, ..., mag\_z\_delve} & AB Kron magnitudes taken from DELVE DR2 by default\\
\texttt{msgerr\_g\_delve, ..., ,magerr\_z\_delve} & Magnitude errors from DELVE DR2 \\
\texttt{mag\_g\_decals, ..., mag\_z\_decals} & AB Kron magnitudes taken from DECaLS DR10 photometry \\
\texttt{snr\_g\_decals, ..., snr\_z\_decals} & Signal-to-noise ratios from DECaLS DR10 \\
\texttt{ypetromag, ..., kspetromag} & YJHKs Petrosian magnitudes from VHS \\
\texttt{ypetromagerr, ..., kspetromagerr} & YJHKs magnitude errors from VHS \\
\texttt{w1mpro, w2mpro}, \texttt{w1sigmpro, w2sigmpro} & WISE W1 and W2 magnitudes and errors \\
\texttt{BD\_chi2\_min}, \texttt{BD\_chi2\_template} & $\chi^2_{\text{BD,min}}$ for brown dwarf templates and best-fit BD template type \\
\texttt{QSO\_chi2\_min}, \texttt{QSO\_z} & $\chi^2_{\text{QSO,min}}$ value for quasar templates and photometric redshifts \\
\texttt{R\_chi2} & QSO to BD $\chi^2$ ratio; $<1$ means QSO fit is better \\
\texttt{F\_test\_value} & F-test values \\
\texttt{BIC\_value} & Bayesian Information Criterion values \\
\texttt{QSO\_EBV}, \texttt{QSO\_EMline}, \texttt{QSO\_Par} & QSO template parameters \\
\texttt{Data\_Points} & Number of bands used in the fit \\
\texttt{ebv} & Galactic $E(B-V)$ from SFD maps \\
\texttt{Prioritization} & Prioritization of the sources based on $\chi^2_{\text{QSO,min}}$ and $R_{\chi^2}$ \\
\hline
\end{tabular}
\end{table*}

\begin{acknowledgements}
We thank the anonymous referee for their valuable suggestions, which improved the quality of this paper. We also thank the journal editor for their time, effort, and careful coordination of the refereeing process. We acknowledge support from Fondecyt Iniciación grant 11240336 (TM, CM), the ANID BASAL project FB210003 (TM, CM, RJA, ARL, FEB, LNMR, SSS), Fondecyt Regular grant 1231718 (RJA), Fondecyt Regular grant 1241005 (FEB, LNMR), Millennium Science Initiative, AIM23-0001 and ICN12\_009 (FEB, LNMR), the National Agency for Research and Development (ANID) under the fellowship ANID Becas/Doctorado Nacional\#21220337 (LNMR-R) and the Vicerrectoría de Investigación of Pontificia Universidad Católica de Chile under the fellowship Stay of Doctoral Co-tutelage Abroad, leading to double degree (LNM-R). MJT acknowledges funding from a FONDECYT Postdoctoral fellowship (3220516) and from UKRI STFC (ST/X001075/1). The work of DS was carried out at the Jet Propulsion Laboratory, California Institute of Technology, under a contract with the National Aeronautics and Space Administration (80NM0018D0004). E.P.F. is supported by the international Gemini Observatory, a program of NSF NOIRLab, which is managed by the Association of Universities for Research in Astronomy (AURA) under a cooperative agreement with the U.S. National Science Foundation, on behalf of the Gemini partnership of Argentina, Brazil, Canada, Chile, the Republic of Korea, and the United States of America. This work is part of the 4-meter Multi-Object Spectroscopic Telescope (4MOST) Chilean AGN/Galaxy Extragalactic Survey (ChANGES) collaboration, and we acknowledge the support of the ChANGES team. We are grateful to the ESO staff at La Silla Observatory for their support during the NTT/EFOSC2 observations under program 115.2816.001, and to the Palomar/NGPS observations that confirmed 3 high-redshift quasar candidate. This work is based on data from the Dark Energy Camera Local Volume Exploration Survey (DELVE DR2), the Dark Energy Camera Legacy Survey (DECaLS DR10), the VISTA Hemisphere Survey (VHS DR5), CatWISE 2020 catalog and Gaia DR3. We acknowledge using data products from the NOIRLab Datalab.
\end{acknowledgements}
\clearpage

\bibliographystyle{aa} 
\bibliography{references} 

\begin{thebibliography}{103}
\expandafter\ifx\csname natexlab\endcsname\relax\def\natexlab#1{#1}\fi

\bibitem[{{Abbott} {et~al.}(2018){Abbott}, {Abdalla}, {Allam}, {Amara}, {Annis}, {Asorey}, {Avila}, {Ballester}, {Banerji}, {Barkhouse}, {Baruah}, {Baumer}, {Bechtol}, {Becker}, {Benoit-L{\'e}vy}, {Bernstein}, {Bertin}, {Blazek}, {Bocquet}, {Brooks}, {Brout}, {Buckley-Geer}, {Burke}, {Busti}, {Campisano}, {Cardiel-Sas}, {Carnero Rosell}, {Carrasco Kind}, {Carretero}, {Castander}, {Cawthon}, {Chang}, {Chen}, {Conselice}, {Costa}, {Crocce}, {Cunha}, {D'Andrea}, {da Costa}, {Das}, {Daues}, {Davis}, {Davis}, {De Vicente}, {DePoy}, {DeRose}, {Desai}, {Diehl}, {Dietrich}, {Dodelson}, {Doel}, {Drlica-Wagner}, {Eifler}, {Elliott}, {Evrard}, {Farahi}, {Fausti Neto}, {Fernandez}, {Finley}, {Flaugher}, {Foley}, {Fosalba}, {Friedel}, {Frieman}, {Garc{\'\i}a-Bellido}, {Gaztanaga}, {Gerdes}, {Giannantonio}, {Gill}, {Glazebrook}, {Goldstein}, {Gower}, {Gruen}, {Gruendl}, {Gschwend}, {Gupta}, {Gutierrez}, {Hamilton}, {Hartley}, {Hinton}, {Hislop}, {Hollowood}, {Honscheid}, {Hoyle}, {Huterer}, {Jain}, {James}, {Jeltema},
  {Johnson}, {Johnson}, {Kacprzak}, {Kent}, {Khullar}, {Klein}, {Kovacs}, {Koziol}, {Krause}, {Kremin}, {Kron}, {Kuehn}, {Kuhlmann}, {Kuropatkin}, {Lahav}, {Lasker}, {Li}, {Li}, {Liddle}, {Lima}, {Lin}, {L{\'o}pez-Reyes}, {MacCrann}, {Maia}, {Maloney}, {Manera}, {March}, {Marriner}, {Marshall}, {Martini}, {McClintock}, {McKay}, {McMahon}, {Melchior}, {Menanteau}, {Miller}, {Miquel}, {Mohr}, {Morganson}, {Mould}, {Neilsen}, {Nichol}, {Nogueira}, {Nord}, {Nugent}, {Nunes}, {Ogando}, {Old}, {Pace}, {Palmese}, {Paz-Chinch{\'o}n}, {Peiris}, {Percival}, {Petravick}, {Plazas}, {Poh}, {Pond}, {Porredon}, {Pujol}, {Refregier}, {Reil}, {Ricker}, {Rollins}, {Romer}, {Roodman}, {Rooney}, {Ross}, {Rykoff}, {Sako}, {Sanchez}, {Sanchez}, {Santiago}, {Saro}, {Scarpine}, {Scolnic}, {Serrano}, {Sevilla-Noarbe}, {Sheldon}, {Shipp}, {Silveira}, {Smith}, {Smith}, {Smith}, {Soares-Santos}, {Sobreira}, {Song}, {Stebbins}, {Suchyta}, {Sullivan}, {Swanson}, {Tarle}, {Thaler}, {Thomas}, {Thomas}, {Troxel}, {Tucker}, {Vikram}, {Vivas},
  {Walker}, {Wechsler}, {Weller}, {Wester}, {Wolf}, {Wu}, {Yanny}, {Zenteno}, {Zhang}, {Zuntz}, {DES Collaboration}, {Juneau}, {Fitzpatrick}, \& {Nikutta}}]{Abbott_2018}
{Abbott}, T.~M.~C., {Abdalla}, F.~B., {Allam}, S., {et~al.} 2018, \apjs, 239, 18

\bibitem[{{Ahumada} {et~al.}(2020){Ahumada}, {Allende Prieto}, {Almeida}, {Anders}, {Anderson}, {Andrews}, {Anguiano}, {Arcodia}, {Armengaud}, {Aubert}, {Avila}, {Avila-Reese}, {Badenes}, {Balland}, {Barger}, {Barrera-Ballesteros}, {Basu}, {Bautista}, {Beaton}, {Beers}, {Benavides}, {Bender}, {Bernardi}, {Bershady}, {Beutler}, {Bidin}, {Bird}, {Bizyaev}, {Blanc}, {Blanton}, {Boquien}, {Borissova}, {Bovy}, {Brandt}, {Brinkmann}, {Brownstein}, {Bundy}, {Bureau}, {Burgasser}, {Burtin}, {Cano-D{\'\i}az}, {Capasso}, {Cappellari}, {Carrera}, {Chabanier}, {Chaplin}, {Chapman}, {Cherinka}, {Chiappini}, {Doohyun Choi}, {Chojnowski}, {Chung}, {Clerc}, {Coffey}, {Comerford}, {Comparat}, {da Costa}, {Cousinou}, {Covey}, {Crane}, {Cunha}, {Ilha}, {Dai}, {Damsted}, {Darling}, {Davidson}, {Davies}, {Dawson}, {De}, {de la Macorra}, {De Lee}, {Queiroz}, {Deconto Machado}, {de la Torre}, {Dell'Agli}, {du Mas des Bourboux}, {Diamond-Stanic}, {Dillon}, {Donor}, {Drory}, {Duckworth}, {Dwelly}, {Ebelke}, {Eftekharzadeh}, {Davis
  Eigenbrot}, {Elsworth}, {Eracleous}, {Erfanianfar}, {Escoffier}, {Fan}, {Farr}, {Fern{\'a}ndez-Trincado}, {Feuillet}, {Finoguenov}, {Fofie}, {Fraser-McKelvie}, {Frinchaboy}, {Fromenteau}, {Fu}, {Galbany}, {Garcia}, {Garc{\'\i}a-Hern{\'a}ndez}, {Garma Oehmichen}, {Ge}, {Geimba Maia}, {Geisler}, {Gelfand}, {Goddy}, {Gonzalez-Perez}, {Grabowski}, {Green}, {Grier}, {Guo}, {Guy}, {Harding}, {Hasselquist}, {Hawken}, {Hayes}, {Hearty}, {Hekker}, {Hogg}, {Holtzman}, {Horta}, {Hou}, {Hsieh}, {Huber}, {Hunt}, {Ider Chitham}, {Imig}, {Jaber}, {Jimenez Angel}, {Johnson}, {Jones}, {J{\"o}nsson}, {Jullo}, {Kim}, {Kinemuchi}, {Kirkpatrick}, {Kite}, {Klaene}, {Kneib}, {Kollmeier}, {Kong}, {Kounkel}, {Krishnarao}, {Lacerna}, {Lan}, {Lane}, {Law}, {Le Goff}, {Leung}, {Lewis}, {Li}, {Lian}, {Lin}, {Long}, {Longa-Pe{\~n}a}, {Lundgren}, {Lyke}, {Mackereth}, {MacLeod}, {Majewski}, {Manchado}, {Maraston}, {Martini}, {Masseron}, {Masters}, {Mathur}, {McDermid}, {Merloni}, {Merrifield}, {M{\'e}sz{\'a}ros}, {Miglio}, {Minniti},
  {Minsley}, {Miyaji}, {Mohammad}, {Mosser}, {Mueller}, {Muna}, {Mu{\~n}oz-Guti{\'e}rrez}, {Myers}, {Nadathur}, {Nair}, {Nandra}, {Correa do Nascimento}, {Nevin}, {Newman}, {Nidever}, {Nitschelm}, {Noterdaeme}, {O'Connell}, {Olmstead}, {Oravetz}, {Oravetz}, {Osorio}, {Pace}, {Padilla}, {Palanque-Delabrouille}, \& {Palicio}}]{Ahumada_2020}
{Ahumada}, R., {Allende Prieto}, C., {Almeida}, A., {et~al.} 2020, \apjs, 249, 3

\bibitem[{{Allard} {et~al.}(2007){Allard}, {Allard}, {Homeier}, {Kielkopf}, {McCaughrean}, \& {Spiegelman}}]{Allard_2007}
{Allard}, F., {Allard}, N.~F., {Homeier}, D., {et~al.} 2007, \aap, 474, L21

\bibitem[{{Assef} {et~al.}(2018){Assef}, {Stern}, {Noirot}, {Jun}, {Cutri}, \& {Eisenhardt}}]{Assef_2018}
{Assef}, R.~J., {Stern}, D., {Noirot}, G., {et~al.} 2018, \apjs, 234, 23

\bibitem[{{Ba{\~n}ados} {et~al.}(2018){Ba{\~n}ados}, {Carilli}, {Walter}, {Momjian}, {Decarli}, {Farina}, {Mazzucchelli}, \& {Venemans}}]{Banados_2018}
{Ba{\~n}ados}, E., {Carilli}, C., {Walter}, F., {et~al.} 2018, \apjl, 861, L14

\bibitem[{{Ba{\~n}ados} {et~al.}(2023){Ba{\~n}ados}, {Schindler}, {Venemans}, {Connor}, {Decarli}, {Farina}, {Mazzucchelli}, {Meyer}, {Stern}, {Walter}, {Fan}, {Hennawi}, {Khusanova}, {Morrell}, {Nanni}, {Noirot}, {Pensabene}, {Rix}, {Simon}, {Verdoes Kleijn}, {Xie}, {Yang}, \& {Connor}}]{Banados_2023}
{Ba{\~n}ados}, E., {Schindler}, J.-T., {Venemans}, B.~P., {et~al.} 2023, \apjs, 265, 29

\bibitem[{{Ba{\~n}ados} {et~al.}(2016){Ba{\~n}ados}, {Venemans}, {Decarli}, {Farina}, {Mazzucchelli}, {Walter}, {Fan}, {Stern}, {Schlafly}, {Chambers}, {Rix}, {Jiang}, {McGreer}, {Simcoe}, {Wang}, {Yang}, {Morganson}, {De Rosa}, {Greiner}, {Balokovi{\'c}}, {Burgett}, {Cooper}, {Draper}, {Flewelling}, {Hodapp}, {Jun}, {Kaiser}, {Kudritzki}, {Magnier}, {Metcalfe}, {Miller}, {Schindler}, {Tonry}, {Wainscoat}, {Waters}, \& {Yang}}]{Banados_2016}
{Ba{\~n}ados}, E., {Venemans}, B.~P., {Decarli}, R., {et~al.} 2016, \apjs, 227, 11

\bibitem[{{Bauer} {et~al.}(2023){Bauer}, {Lira}, {Anguita}, {Arevalo}, {Assef}, {Barrientos}, {Berg}, {Bernal}, {Bian}, {Boquien}, {Buat}, {Chilingarian}, {Coppi}, {De Cicco}, {Diaz}, {Grishin}, {Hernandez-Garcia}, {Kakkad}, {Katkov}, {Krogager}, {L{\'o}pez-Navas}, {Mart{\'\i}nez-Ram{\'\i}rez}, {Mazzucchelli}, {Motta}, {Ricci}, {Ricci}, {Rojas}, {Rouse}, {S{\'a}nchez-S{\'a}ez}, {Toptun}, {Treister}, \& {Vito}}]{Bauer_2023}
{Bauer}, F.~E., {Lira}, P., {Anguita}, T., {et~al.} 2023, The Messenger, 190, 34

\bibitem[{{Becker} {et~al.}(2015){Becker}, {Bolton}, \& {Lidz}}]{Becker_2015}
{Becker}, G.~D., {Bolton}, J.~S., \& {Lidz}, A. 2015, \pasa, 32, e045

\bibitem[{{Belladitta} {et~al.}(2025){Belladitta}, {Ba{\~n}ados}, {Xie}, {Decarli}, {Onorato}, {Yang}, {Bischetti}, {Onoue}, {Loiacono}, {Mart{\'\i}nez-Ram{\'\i}rez}, {Mazzucchelli}, {Davies}, {Wolf}, {Schindler}, {Fan}, {Wang}, {Walter}, {Mkrtchyan}, {Stern}, {Farina}, \& {Venemans}}]{Belladitta_2025}
{Belladitta}, S., {Ba{\~n}ados}, E., {Xie}, Z.-L., {et~al.} 2025, \aap, 699, A335

\bibitem[{{Belladitta} {et~al.}(2023){Belladitta}, {Moretti}, {Caccianiga}, {Dallacasa}, {Spingola}, {Pedani}, {Cassar{\`a}}, \& {Bisogni}}]{Belladitta_2023}
{Belladitta}, S., {Moretti}, A., {Caccianiga}, A., {et~al.} 2023, \aap, 669, A134

\bibitem[{{Belladitta} {et~al.}(2019){Belladitta}, {Moretti}, {Caccianiga}, {Ghisellini}, {Cicone}, {Sbarrato}, {Ighina}, \& {Pedani}}]{Belladitta_2019}
{Belladitta}, S., {Moretti}, A., {Caccianiga}, A., {et~al.} 2019, \aap, 629, A68

\bibitem[{{Bertin} \& {Arnouts}(1996)}]{Bertin_1996}
{Bertin}, E. \& {Arnouts}, S. 1996, \aaps, 117, 393

\bibitem[{{Best} {et~al.}(2015){Best}, {Liu}, {Magnier}, {Deacon}, {Aller}, {Redstone}, {Burgett}, {Chambers}, {Draper}, {Flewelling}, {Hodapp}, {Kaiser}, {Metcalfe}, {Tonry}, {Wainscoat}, \& {Waters}}]{Best_2015}
{Best}, W. M.~J., {Liu}, M.~C., {Magnier}, E.~A., {et~al.} 2015, \apj, 814, 118

\bibitem[{{Bogd{\'a}n} {et~al.}(2024){Bogd{\'a}n}, {Goulding}, {Natarajan}, {Kov{\'a}cs}, {Tremblay}, {Chadayammuri}, {Volonteri}, {Kraft}, {Forman}, {Jones}, {Churazov}, \& {Zhuravleva}}]{Bogdan_2024}
{Bogd{\'a}n}, {\'A}., {Goulding}, A.~D., {Natarajan}, P., {et~al.} 2024, Nature Astronomy, 8, 126

\bibitem[{{Bosman} {et~al.}(2022){Bosman}, {Davies}, {Becker}, {Keating}, {Davies}, {Zhu}, {Eilers}, {D'Odorico}, {Bian}, {Bischetti}, {Cristiani}, {Fan}, {Farina}, {Haehnelt}, {Hennawi}, {Kulkarni}, {Mesinger}, {Meyer}, {Onoue}, {Pallottini}, {Qin}, {Ryan-Weber}, {Schindler}, {Walter}, {Wang}, \& {Yang}}]{Bosman_2022}
{Bosman}, S. E.~I., {Davies}, F.~B., {Becker}, G.~D., {et~al.} 2022, \mnras, 514, 55

\bibitem[{{Brandt} {et~al.}(2001){Brandt}, {Alexander}, {Hornschemeier}, {Garmire}, {Schneider}, {Barger}, {Bauer}, {Broos}, {Cowie}, {Townsley}, {Burrows}, {Chartas}, {Feigelson}, {Griffiths}, {Nousek}, \& {Sargent}}]{Brandt_2001}
{Brandt}, W.~N., {Alexander}, D.~M., {Hornschemeier}, A.~E., {et~al.} 2001, \aj, 122, 2810

\bibitem[{{Buzzoni} {et~al.}(1984){Buzzoni}, {Delabre}, {Dekker}, {Dodorico}, {Enard}, {Focardi}, {Gustafsson}, {Nees}, {Paureau}, \& {Reiss}}]{Buzzoni_1984}
{Buzzoni}, B., {Delabre}, B., {Dekker}, H., {et~al.} 1984, The Messenger, 38, 9

\bibitem[{{Byrne} {et~al.}(2024){Byrne}, {Meyer}, {Farina}, {Ba{\~n}ados}, {Walter}, {Decarli}, {Belladitta}, \& {Loiacono}}]{Byrne_2024}
{Byrne}, X., {Meyer}, R.~A., {Farina}, E.~P., {et~al.} 2024, \mnras, 530, 870

\bibitem[{{Calderone} {et~al.}(2024){Calderone}, {Guarneri}, {Porru}, {Cristiani}, {Grazian}, {Nicastro}, {Bischetti}, {Boutsia}, {Cupani}, {D'Odorico}, {Feruglio}, \& {Fontanot}}]{Calderone_2024}
{Calderone}, G., {Guarneri}, F., {Porru}, M., {et~al.} 2024, \aap, 683, A34

\bibitem[{{Chambers} {et~al.}(2016){Chambers}, {Magnier}, {Metcalfe}, {Flewelling}, {Huber}, {Waters}, {Denneau}, {Draper}, {Farrow}, {Finkbeiner}, {Holmberg}, {Koppenhoefer}, {Price}, {Rest}, {Saglia}, {Schlafly}, {Smartt}, {Sweeney}, {Wainscoat}, {Burgett}, {Chastel}, {Grav}, {Heasley}, {Hodapp}, {Jedicke}, {Kaiser}, {Kudritzki}, {Luppino}, {Lupton}, {Monet}, {Morgan}, {Onaka}, {Shiao}, {Stubbs}, {Tonry}, {White}, {Ba{\~n}ados}, {Bell}, {Bender}, {Bernard}, {Boegner}, {Boffi}, {Botticella}, {Calamida}, {Casertano}, {Chen}, {Chen}, {Cole}, {Deacon}, {Frenk}, {Fitzsimmons}, {Gezari}, {Gibbs}, {Goessl}, {Goggia}, {Gourgue}, {Goldman}, {Grant}, {Grebel}, {Hambly}, {Hasinger}, {Heavens}, {Heckman}, {Henderson}, {Henning}, {Holman}, {Hopp}, {Ip}, {Isani}, {Jackson}, {Keyes}, {Koekemoer}, {Kotak}, {Le}, {Liska}, {Long}, {Lucey}, {Liu}, {Martin}, {Masci}, {McLean}, {Mindel}, {Misra}, {Morganson}, {Murphy}, {Obaika}, {Narayan}, {Nieto-Santisteban}, {Norberg}, {Peacock}, {Pier}, {Postman}, {Primak}, {Rae}, {Rai},
  {Riess}, {Riffeser}, {Rix}, {R{\"o}ser}, {Russel}, {Rutz}, {Schilbach}, {Schultz}, {Scolnic}, {Strolger}, {Szalay}, {Seitz}, {Small}, {Smith}, {Soderblom}, {Taylor}, {Thomson}, {Taylor}, {Thakar}, {Thiel}, {Thilker}, {Unger}, {Urata}, {Valenti}, {Wagner}, {Walder}, {Walter}, {Watters}, {Werner}, {Wood-Vasey}, \& {Wyse}}]{Chambers_2016}
{Chambers}, K.~C., {Magnier}, E.~A., {Metcalfe}, N., {et~al.} 2016, arXiv e-prints, arXiv:1612.05560

\bibitem[{{Champagne} {et~al.}(2025{\natexlab{a}}){Champagne}, {Wang}, {Yang}, {Fan}, {Hennawi}, {Sun}, {Ba{\~n}ados}, {Bosman}, {Costa}, {Habouzit}, {Jin}, {Jun}, {Li}, {Liu}, {Loiacono}, {Lupi}, {Mazzucchelli}, {Pudoka}, {Rojas-Ruiz}, {Tee}, {Trebitsch}, {Zhang}, {Zhuang}, \& {Zou}}]{Champagne_2025a}
{Champagne}, J.~B., {Wang}, F., {Yang}, J., {et~al.} 2025{\natexlab{a}}, \apj, 981, 114

\bibitem[{{Champagne} {et~al.}(2025{\natexlab{b}}){Champagne}, {Wang}, {Zhang}, {Yang}, {Fan}, {Hennawi}, {Sun}, {Ba{\~n}ados}, {Bosman}, {Costa}, {Eilers}, {Endsley}, {Jin}, {Jun}, {Li}, {Lin}, {Liu}, {Loiacono}, {Lupi}, {Mazzucchelli}, {Pudoka}, {Protu{\v{s}}ov{\`a}}, {Rojas-Ruiz}, {Tee}, {Trebitsch}, {Venemans}, {Zhuang}, \& {Zou}}]{Champagne_2025b}
{Champagne}, J.~B., {Wang}, F., {Zhang}, H., {et~al.} 2025{\natexlab{b}}, \apj, 981, 113

\bibitem[{{Chiu} {et~al.}(2007){Chiu}, {Richards}, {Hewett}, \& {Maddox}}]{Chiu_2007}
{Chiu}, K., {Richards}, G.~T., {Hewett}, P.~C., \& {Maddox}, N. 2007, \mnras, 375, 1180

\bibitem[{{Chung} {et~al.}(2014){Chung}, {Kochanek}, {Assef}, {Brown}, {Stern}, {Jannuzi}, {Gonzalez}, {Hickox}, \& {Moustakas}}]{Chung_2014}
{Chung}, S.~M., {Kochanek}, C.~S., {Assef}, R., {et~al.} 2014, \apj, 790, 54

\bibitem[{{Davies} {et~al.}(2023){Davies}, {Ryan-Weber}, {D'Odorico}, {Bosman}, {Meyer}, {Becker}, {Cupani}, {Bischetti}, {Sebastian}, {Eilers}, {Farina}, {Wang}, {Yang}, \& {Zhu}}]{Davies_2023}
{Davies}, R.~L., {Ryan-Weber}, E., {D'Odorico}, V., {et~al.} 2023, \mnras, 521, 289

\bibitem[{{de Jong} {et~al.}(2019){de Jong}, {Agertz}, {Berbel}, {Aird}, {Alexander}, {Amarsi}, {Anders}, {Andrae}, {Ansarinejad}, {Ansorge}, {Antilogus}, {Anwand-Heerwart}, {Arentsen}, {Arnadottir}, {Asplund}, {Auger}, {Azais}, {Baade}, {Baker}, {Baker}, {Balbinot}, {Baldry}, {Banerji}, {Barden}, {Barklem}, {Barth{\'e}l{\'e}my-Mazot}, {Battistini}, {Bauer}, {Bell}, {Bellido-Tirado}, {Bellstedt}, {Belokurov}, {Bensby}, {Bergemann}, {Bestenlehner}, {Bielby}, {Bilicki}, {Blake}, {Bland-Hawthorn}, {Boeche}, {Boland}, {Boller}, {Bongard}, {Bongiorno}, {Bonifacio}, {Boudon}, {Brooks}, {Brown}, {Brown}, {Br{\"u}ggen}, {Brynnel}, {Brzeski}, {Buchert}, {Buschkamp}, {Caffau}, {Caillier}, {Carrick}, {Casagrande}, {Case}, {Casey}, {Cesarini}, {Cescutti}, {Chapuis}, {Chiappini}, {Childress}, {Christlieb}, {Church}, {Cioni}, {Cluver}, {Colless}, {Collett}, {Comparat}, {Cooper}, {Couch}, {Courbin}, {Croom}, {Croton}, {Daguis{\'e}}, {Dalton}, {Davies}, {Davis}, {de Laverny}, {Deason}, {Dionies}, {Disseau}, {Doel},
  {D{\"o}scher}, {Driver}, {Dwelly}, {Eckert}, {Edge}, {Edvardsson}, {Youssoufi}, {Elhaddad}, {Enke}, {Erfanianfar}, {Farrell}, {Fechner}, {Feiz}, {Feltzing}, {Ferreras}, {Feuerstein}, {Feuillet}, {Finoguenov}, {Ford}, {Fotopoulou}, {Fouesneau}, {Frenk}, {Frey}, {Gaessler}, {Geier}, {Gentile Fusillo}, {Gerhard}, {Giannantonio}, {Giannone}, {Gibson}, {Gillingham}, {Gonz{\'a}lez-Fern{\'a}ndez}, {Gonzalez-Solares}, {Gottloeber}, {Gould}, {Grebel}, {Gueguen}, {Guiglion}, {Haehnelt}, {Hahn}, {Hansen}, {Hartman}, {Hauptner}, {Hawkins}, {Haynes}, {Haynes}, {Heiter}, {Helmi}, {Aguayo}, {Hewett}, {Hinton}, {Hobbs}, {Hoenig}, {Hofman}, {Hook}, {Hopgood}, {Hopkins}, {Hourihane}, {Howes}, {Howlett}, {Huet}, {Irwin}, {Iwert}, {Jablonka}, {Jahn}, {Jahnke}, {Jarno}, {Jin}, {Jofre}, {Johl}, {Jones}, {J{\"o}nsson}, {Jordan}, {Karovicova}, {Khalatyan}, {Kelz}, {Kennicutt}, {King}, {Kitaura}, {Klar}, {Klauser}, {Kneib}, {Koch}, {Koposov}, {Kordopatis}, {Korn}, {Kosmalski}, {Kotak}, {Kovalev}, {Kreckel}, {Kripak}, {Krumpe},
  {Kuijken}, {Kunder}, {Kushniruk}, {Lam}, {Lamer}, {Laurent}, {Lawrence}, {Lehmitz}, {Lemasle}, {Lewis}, {Li}, {Lidman}, {Lind}, {Liske}, {Lizon}, {Loveday}, {Ludwig}, {McDermid}, {Maguire}, {Mainieri}, {Mali}, \& {Mandel}}]{Jong_2019}
{de Jong}, R.~S., {Agertz}, O., {Berbel}, A.~A., {et~al.} 2019, The Messenger

\bibitem[{{Dey} {et~al.}(2019){Dey}, {Schlegel}, {Lang}, {Blum}, {Burleigh}, {Fan}, {Findlay}, {Finkbeiner}, {Herrera}, {Juneau}, {Landriau}, {Levi}, {McGreer}, {Meisner}, {Myers}, {Moustakas}, {Nugent}, {Patej}, {Schlafly}, {Walker}, {Valdes}, {Weaver}, {Y{\`e}che}, {Zou}, {Zhou}, {Abareshi}, {Abbott}, {Abolfathi}, {Aguilera}, {Alam}, {Allen}, {Alvarez}, {Annis}, {Ansarinejad}, {Aubert}, {Beechert}, {Bell}, {BenZvi}, {Beutler}, {Bielby}, {Bolton}, {Brice{\~n}o}, {Buckley-Geer}, {Butler}, {Calamida}, {Carlberg}, {Carter}, {Casas}, {Castander}, {Choi}, {Comparat}, {Cukanovaite}, {Delubac}, {DeVries}, {Dey}, {Dhungana}, {Dickinson}, {Ding}, {Donaldson}, {Duan}, {Duckworth}, {Eftekharzadeh}, {Eisenstein}, {Etourneau}, {Fagrelius}, {Farihi}, {Fitzpatrick}, {Font-Ribera}, {Fulmer}, {G{\"a}nsicke}, {Gaztanaga}, {George}, {Gerdes}, {Gontcho}, {Gorgoni}, {Green}, {Guy}, {Harmer}, {Hernandez}, {Honscheid}, {Huang}, {James}, {Jannuzi}, {Jiang}, {Joyce}, {Karcher}, {Karkar}, {Kehoe}, {Kneib}, {Kueter-Young}, {Lan},
  {Lauer}, {Le Guillou}, {Le Van Suu}, {Lee}, {Lesser}, {Perreault Levasseur}, {Li}, {Mann}, {Marshall}, {Mart{\'\i}nez-V{\'a}zquez}, {Martini}, {du Mas des Bourboux}, {McManus}, {Meier}, {M{\'e}nard}, {Metcalfe}, {Mu{\~n}oz-Guti{\'e}rrez}, {Najita}, {Napier}, {Narayan}, {Newman}, {Nie}, {Nord}, {Norman}, {Olsen}, {Paat}, {Palanque-Delabrouille}, {Peng}, {Poppett}, {Poremba}, {Prakash}, {Rabinowitz}, {Raichoor}, {Rezaie}, {Robertson}, {Roe}, {Ross}, {Ross}, {Rudnick}, {Safonova}, {Saha}, {S{\'a}nchez}, {Savary}, {Schweiker}, {Scott}, {Seo}, {Shan}, {Silva}, {Slepian}, {Soto}, {Sprayberry}, {Staten}, {Stillman}, {Stupak}, {Summers}, {Sien Tie}, {Tirado}, {Vargas-Maga{\~n}a}, {Vivas}, {Wechsler}, {Williams}, {Yang}, {Yang}, {Yapici}, {Zaritsky}, {Zenteno}, {Zhang}, {Zhang}, {Zhou}, \& {Zhou}}]{Dey_2019}
{Dey}, A., {Schlegel}, D.~J., {Lang}, D., {et~al.} 2019, \aj, 157, 168

\bibitem[{{DiPompeo} {et~al.}(2015){DiPompeo}, {Myers}, {Hickox}, {Geach}, {Holder}, {Hainline}, \& {Hall}}]{DiPompeo_2015}
{DiPompeo}, M.~A., {Myers}, A.~D., {Hickox}, R.~C., {et~al.} 2015, \mnras, 446, 3492

\bibitem[{{Drlica-Wagner} {et~al.}(2022){Drlica-Wagner}, {Ferguson}, {Adam{\'o}w}, {Aguena}, {Allam}, {Andrade-Oliveira}, {Bacon}, {Bechtol}, {Bell}, {Bertin}, {Bilaji}, {Bocquet}, {Bom}, {Brooks}, {Burke}, {Carballo-Bello}, {Carlin}, {Carnero Rosell}, {Carrasco Kind}, {Carretero}, {Castander}, {Cerny}, {Chang}, {Choi}, {Conselice}, {Costanzi}, {Crnojevi{\'c}}, {da Costa}, {de Vicente}, {Desai}, {Esteves}, {Everett}, {Ferrero}, {Fitzpatrick}, {Flaugher}, {Friedel}, {Frieman}, {Garc{\'\i}a-Bellido}, {Gatti}, {Gaztanaga}, {Gerdes}, {Gruen}, {Gruendl}, {Gschwend}, {Hartley}, {Hernandez-Lang}, {Hinton}, {Hollowood}, {Honscheid}, {Hughes}, {Jacques}, {James}, {Johnson}, {Kuehn}, {Kuropatkin}, {Lahav}, {Li}, {Lidman}, {Lin}, {March}, {Marshall}, {Mart{\'\i}nez-Delgado}, {Mart{\'\i}nez-V{\'a}zquez}, {Massana}, {Mau}, {McNanna}, {Melchior}, {Menanteau}, {Miller}, {Miquel}, {Mohr}, {Morgan}, {Mutlu-Pakdil}, {Mu{\~n}oz}, {Neilsen}, {Nidever}, {Nikutta}, {Nilo Castellon}, {No{\"e}l}, {Ogando}, {Olsen}, {Pace},
  {Palmese}, {Paz-Chinch{\'o}n}, {Pereira}, {Pieres}, {Plazas Malag{\'o}n}, {Prat}, {Riley}, {Rodriguez-Monroy}, {Romer}, {Roodman}, {Sako}, {Sakowska}, {Sanchez}, {S{\'a}nchez}, {Sand}, {Santana-Silva}, {Santiago}, {Schubnell}, {Serrano}, {Sevilla-Noarbe}, {Simon}, {Smith}, {Soares-Santos}, {Stringfellow}, {Suchyta}, {Suson}, {Tan}, {Tarle}, {Tavangar}, {Thomas}, {To}, {Tollerud}, {Troxel}, {Tucker}, {Varga}, {Vivas}, {Walker}, {Weller}, {Wilkinson}, {Wu}, {Yanny}, {Zaborowski}, {Zenteno}, {Delve Collaboration}, {Des Collaboration}, \& {Astro Data Lab}}]{Drlica_Wagner_2022}
{Drlica-Wagner}, A., {Ferguson}, P.~S., {Adam{\'o}w}, M., {et~al.} 2022, \apjs, 261, 38

\bibitem[{{Edge} {et~al.}(2013){Edge}, {Sutherland}, {Kuijken}, {Driver}, {McMahon}, {Eales}, \& {Emerson}}]{Edge_2013}
{Edge}, A., {Sutherland}, W., {Kuijken}, K., {et~al.} 2013, The Messenger, 154, 32

\bibitem[{{Euclid Collaboration} {et~al.}(2019){Euclid Collaboration}, {Barnett}, {Warren}, {Mortlock}, {Cuby}, {Conselice}, {Hewett}, {Willott}, {Auricchio}, {Balaguera-Antol{\'\i}nez}, {Baldi}, {Bardelli}, {Bellagamba}, {Bender}, {Biviano}, {Bonino}, {Bozzo}, {Branchini}, {Brescia}, {Brinchmann}, {Burigana}, {Camera}, {Capobianco}, {Carbone}, {Carretero}, {Carvalho}, {Castander}, {Castellano}, {Cavuoti}, {Cimatti}, {Cl{\'e}dassou}, {Congedo}, {Conversi}, {Copin}, {Corcione}, {Coupon}, {Courtois}, {Cropper}, {Da Silva}, {Duncan}, {Dusini}, {Ealet}, {Farrens}, {Fosalba}, {Fotopoulou}, {Fourmanoit}, {Frailis}, {Fumana}, {Galeotta}, {Garilli}, {Gillard}, {Gillis}, {Graci{\'a}-Carpio}, {Grupp}, {Hoekstra}, {Hormuth}, {Israel}, {Jahnke}, {Kermiche}, {Kilbinger}, {Kirkpatrick}, {Kitching}, {Kohley}, {Kubik}, {Kunz}, {Kurki-Suonio}, {Laureijs}, {Ligori}, {Lilje}, {Lloro}, {Maiorano}, {Mansutti}, {Marggraf}, {Martinet}, {Marulli}, {Massey}, {Mauri}, {Medinaceli}, {Mei}, {Mellier}, {Metcalf}, {Metge}, {Meylan},
  {Moresco}, {Moscardini}, {Munari}, {Neissner}, {Niemi}, {Nutma}, {Padilla}, {Paltani}, {Pasian}, {Paykari}, {Percival}, {Pettorino}, {Polenta}, {Poncet}, {Pozzetti}, {Raison}, {Renzi}, {Rhodes}, {Rix}, {Romelli}, {Roncarelli}, {Rossetti}, {Saglia}, {Sapone}, {Scaramella}, {Schneider}, {Scottez}, {Secroun}, {Serrano}, {Sirri}, {Stanco}, {Sureau}, {Tallada-Cresp{\'\i}}, {Tavagnacco}, {Taylor}, {Tenti}, {Tereno}, {Toledo-Moreo}, {Torradeflot}, {Valenziano}, {Vassallo}, {Wang}, {Zacchei}, {Zamorani}, {Zoubian}, \& {Zucca}}]{Barnett_2019}
{Euclid Collaboration}, {Barnett}, R., {Warren}, S.~J., {et~al.} 2019, \aap, 631, A85

\bibitem[{{Fan} {et~al.}(2023){Fan}, {Ba{\~n}ados}, \& {Simcoe}}]{Fan_2023}
{Fan}, X., {Ba{\~n}ados}, E., \& {Simcoe}, R.~A. 2023, \araa, 61, 373

\bibitem[{{Fan} {et~al.}(2001){Fan}, {Narayanan}, {Lupton}, {Strauss}, {Knapp}, {Becker}, {White}, {Pentericci}, {Leggett}, {Haiman}, {Gunn}, {Ivezi{\'c}}, {Schneider}, {Anderson}, {Brinkmann}, {Bahcall}, {Connolly}, {Csabai}, {Doi}, {Fukugita}, {Geballe}, {Grebel}, {Harbeck}, {Hennessy}, {Lamb}, {Miknaitis}, {Munn}, {Nichol}, {Okamura}, {Pier}, {Prada}, {Richards}, {Szalay}, \& {York}}]{Fan_2001}
{Fan}, X., {Narayanan}, V.~K., {Lupton}, R.~H., {et~al.} 2001, \aj, 122, 2833

\bibitem[{{Farina} {et~al.}(2022){Farina}, {Schindler}, {Walter}, {Ba{\~n}ados}, {Davies}, {Decarli}, {Eilers}, {Fan}, {Hennawi}, {Mazzucchelli}, {Meyer}, {Trakhtenbrot}, {Volonteri}, {Wang}, {Worseck}, {Yang}, {Gutcke}, {Venemans}, {Bosman}, {Costa}, {De Rosa}, {Drake}, \& {Onoue}}]{Farina_2022}
{Farina}, E.~P., {Schindler}, J.-T., {Walter}, F., {et~al.} 2022, \apj, 941, 106

\bibitem[{{Findlay} {et~al.}(2012){Findlay}, {Sutherland}, {Venemans}, {Reyl{\'e}}, {Robin}, {Bonfield}, {Bruce}, \& {Jarvis}}]{Findlay_2012}
{Findlay}, J.~R., {Sutherland}, W.~J., {Venemans}, B.~P., {et~al.} 2012, \mnras, 419, 3354

\bibitem[{{Flaugher}(2005)}]{Flaugher_2005}
{Flaugher}, B. 2005, International Journal of Modern Physics A, 20, 3121

\bibitem[{{Flesch}(2023)}]{Flesch_2023}
{Flesch}, E.~W. 2023, The Open Journal of Astrophysics, 6, 49

\bibitem[{{Gaia Collaboration} {et~al.}(2016){Gaia Collaboration}, {Brown}, {Vallenari}, {Prusti}, {de Bruijne}, {Mignard}, {Drimmel}, {Babusiaux}, {Bailer-Jones}, {Bastian}, {Biermann}, {Evans}, {Eyer}, {Jansen}, {Jordi}, {Katz}, {Klioner}, {Lammers}, {Lindegren}, {Luri}, {O'Mullane}, {Panem}, {Pourbaix}, {Randich}, {Sartoretti}, {Siddiqui}, {Soubiran}, {Valette}, {van Leeuwen}, {Walton}, {Aerts}, {Arenou}, {Cropper}, {H{\o}g}, {Lattanzi}, {Grebel}, {Holland}, {Huc}, {Passot}, {Perryman}, {Bramante}, {Cacciari}, {Casta{\~n}eda}, {Chaoul}, {Cheek}, {De Angeli}, {Fabricius}, {Guerra}, {Hern{\'a}ndez}, {Jean-Antoine-Piccolo}, {Masana}, {Messineo}, {Mowlavi}, {Nienartowicz}, {Ord{\'o}{\~n}ez-Blanco}, {Panuzzo}, {Portell}, {Richards}, {Riello}, {Seabroke}, {Tanga}, {Th{\'e}venin}, {Torra}, {Els}, {Gracia-Abril}, {Comoretto}, {Garcia-Reinaldos}, {Lock}, {Mercier}, {Altmann}, {Andrae}, {Astraatmadja}, {Bellas-Velidis}, {Benson}, {Berthier}, {Blomme}, {Busso}, {Carry}, {Cellino}, {Clementini}, {Cowell}, {Creevey},
  {Cuypers}, {Davidson}, {De Ridder}, {de Torres}, {Delchambre}, {Dell'Oro}, {Ducourant}, {Fr{\'e}mat}, {Garc{\'\i}a-Torres}, {Gosset}, {Halbwachs}, {Hambly}, {Harrison}, {Hauser}, {Hestroffer}, {Hodgkin}, {Huckle}, {Hutton}, {Jasniewicz}, {Jordan}, {Kontizas}, {Korn}, {Lanzafame}, {Manteiga}, {Moitinho}, {Muinonen}, {Osinde}, {Pancino}, {Pauwels}, {Petit}, {Recio-Blanco}, {Robin}, {Sarro}, {Siopis}, {Smith}, {Smith}, {Sozzetti}, {Thuillot}, {van Reeven}, {Viala}, {Abbas}, {Abreu Aramburu}, {Accart}, {Aguado}, {Allan}, {Allasia}, {Altavilla}, {{\'A}lvarez}, {Alves}, {Anderson}, {Andrei}, {Anglada Varela}, {Antiche}, {Antoja}, {Ant{\'o}n}, {Arcay}, {Bach}, {Baker}, {Balaguer-N{\'u}{\~n}ez}, {Barache}, {Barata}, {Barbier}, {Barblan}, {Barrado y Navascu{\'e}s}, {Barros}, {Barstow}, {Becciani}, {Bellazzini}, {Bello Garc{\'\i}a}, {Belokurov}, {Bendjoya}, {Berihuete}, {Bianchi}, {Bienaym{\'e}}, {Billebaud}, {Blagorodnova}, {Blanco-Cuaresma}, {Boch}, {Bombrun}, {Borrachero}, {Bouquillon}, {Bourda}, {Bouy},
  {Bragaglia}, {Breddels}, {Brouillet}, {Br{\"u}semeister}, {Bucciarelli}, {Burgess}, {Burgon}, {Burlacu}, {Busonero}, {Buzzi}, {Caffau}, {Cambras}, {Campbell}, {Cancelliere}, {Cantat-Gaudin}, {Carlucci}, {Carrasco}, {Castellani}, {Charlot}, {Charnas}, {Chiavassa}, {Clotet}, {Cocozza}, {Collins}, {Costigan}, {Crifo}, {Cross}, {Crosta}, {Crowley}, {Dafonte}, {Damerdji}, {Dapergolas}, {David}, {David}, \& {De Cat}}]{Brown_2016}
{Gaia Collaboration}, {Brown}, A.~G.~A., {Vallenari}, A., {et~al.} 2016, \aap, 595, A2

\bibitem[{{Gaia Collaboration} {et~al.}(2023){Gaia Collaboration}, {Vallenari}, {Brown}, {Prusti}, {de Bruijne}, {Arenou}, {Babusiaux}, {Biermann}, {Creevey}, {Ducourant}, {Evans}, {Eyer}, {Guerra}, {Hutton}, {Jordi}, {Klioner}, {Lammers}, {Lindegren}, {Luri}, {Mignard}, {Panem}, {Pourbaix}, {Randich}, {Sartoretti}, {Soubiran}, {Tanga}, {Walton}, {Bailer-Jones}, {Bastian}, {Drimmel}, {Jansen}, {Katz}, {Lattanzi}, {van Leeuwen}, {Bakker}, {Cacciari}, {Casta{\~n}eda}, {De Angeli}, {Fabricius}, {Fouesneau}, {Fr{\'e}mat}, {Galluccio}, {Guerrier}, {Heiter}, {Masana}, {Messineo}, {Mowlavi}, {Nicolas}, {Nienartowicz}, {Pailler}, {Panuzzo}, {Riclet}, {Roux}, {Seabroke}, {Sordo}, {Th{\'e}venin}, {Gracia-Abril}, {Portell}, {Teyssier}, {Altmann}, {Andrae}, {Audard}, {Bellas-Velidis}, {Benson}, {Berthier}, {Blomme}, {Burgess}, {Busonero}, {Busso}, {C{\'a}novas}, {Carry}, {Cellino}, {Cheek}, {Clementini}, {Damerdji}, {Davidson}, {de Teodoro}, {Nu{\~n}ez Campos}, {Delchambre}, {Dell'Oro}, {Esquej},
  {Fern{\'a}ndez-Hern{\'a}ndez}, {Fraile}, {Garabato}, {Garc{\'\i}a-Lario}, {Gosset}, {Haigron}, {Halbwachs}, {Hambly}, {Harrison}, {Hern{\'a}ndez}, {Hestroffer}, {Hodgkin}, {Holl}, {Jan{\ss}en}, {Jevardat de Fombelle}, {Jordan}, {Krone-Martins}, {Lanzafame}, {L{\"o}ffler}, {Marchal}, {Marrese}, {Moitinho}, {Muinonen}, {Osborne}, {Pancino}, {Pauwels}, {Recio-Blanco}, {Reyl{\'e}}, {Riello}, {Rimoldini}, {Roegiers}, {Rybizki}, {Sarro}, {Siopis}, {Smith}, {Sozzetti}, {Utrilla}, {van Leeuwen}, {Abbas}, {{\'A}brah{\'a}m}, {Abreu Aramburu}, {Aerts}, {Aguado}, {Ajaj}, {Aldea-Montero}, {Altavilla}, {{\'A}lvarez}, {Alves}, {Anders}, {Anderson}, {Anglada Varela}, {Antoja}, {Baines}, {Baker}, {Balaguer-N{\'u}{\~n}ez}, {Balbinot}, {Balog}, {Barache}, {Barbato}, {Barros}, {Barstow}, {Bartolom{\'e}}, {Bassilana}, {Bauchet}, {Becciani}, {Bellazzini}, {Berihuete}, {Bernet}, {Bertone}, {Bianchi}, {Binnenfeld}, {Blanco-Cuaresma}, {Blazere}, {Boch}, {Bombrun}, {Bossini}, {Bouquillon}, {Bragaglia}, {Bramante}, {Breedt},
  {Bressan}, {Brouillet}, {Brugaletta}, {Bucciarelli}, {Burlacu}, {Butkevich}, {Buzzi}, {Caffau}, {Cancelliere}, {Cantat-Gaudin}, {Carballo}, {Carlucci}, {Carnerero}, {Carrasco}, {Casamiquela}, {Castellani}, {Castro-Ginard}, {Chaoul}, {Charlot}, {Chemin}, {Chiaramida}, {Chiavassa}, {Chornay}, {Comoretto}, {Contursi}, {Cooper}, {Cornez}, {Cowell}, {Crifo}, {Cropper}, {Crosta}, {Crowley}, {Dafonte}, {Dapergolas}, {David}, {David}, {de Laverny}, {De Luise}, \& {De March}}]{Vallenari_2022}
{Gaia Collaboration}, {Vallenari}, A., {Brown}, A.~G.~A., {et~al.} 2023, \aap, 674, A1

\bibitem[{{Giacconi} {et~al.}(2002){Giacconi}, {Zirm}, {Wang}, {Rosati}, {Nonino}, {Tozzi}, {Gilli}, {Mainieri}, {Hasinger}, {Kewley}, {Bergeron}, {Borgani}, {Gilmozzi}, {Grogin}, {Koekemoer}, {Schreier}, {Zheng}, \& {Norman}}]{Giacconi}
{Giacconi}, R., {Zirm}, A., {Wang}, J., {et~al.} 2002, \apjs, 139, 369

\bibitem[{{Gloudemans} {et~al.}(2022){Gloudemans}, {Duncan}, {Saxena}, {Harikane}, {Hill}, {Zeimann}, {R{\"o}ttgering}, {Yang}, {Best}, {Ba{\~n}ados}, {Drabent}, {Hardcastle}, {Hennawi}, {Lansbury}, {Magliocchetti}, {Miley}, {Nanni}, {Shimwell}, {Smith}, {Venemans}, \& {Wagenveld}}]{Gloudemans_2022}
{Gloudemans}, A.~J., {Duncan}, K.~J., {Saxena}, A., {et~al.} 2022, \aap, 668, A27

\bibitem[{{Guiglion} {et~al.}(2019){Guiglion}, {Battistini}, {Bell}, {Bensby}, {Boller}, {Chiappini}, {Comparat}, {Christlieb}, {Church}, {Cioni}, {Davies}, {Dwelly}, {de Jong}, {Feltzing}, {Gueguen}, {Howes}, {Irwin}, {Kushniruk}, {Lam}, {Liske}, {McMahon}, {Merloni}, {Norberg}, {Robotham}, {Schnurr}, {Sorce}, {Starkenburg}, {Storm}, {Swann}, {Tempel}, {Thi}, {Worley}, {Walcher}, \& {4MOST Collaboration}}]{Guiglion_2019}
{Guiglion}, G., {Battistini}, C., {Bell}, C.~P.~M., {et~al.} 2019, The Messenger, 175, 17

\bibitem[{{Hatziminaoglou} {et~al.}(2004){Hatziminaoglou}, {Perez-Fournon}, {Polletta}, {Afonso-Luis}, {Hernan-Caballero}, {Montenegro-Montes}, {Lonsdale}, {Xu}, {Franceschini}, {Rowan-Robinson}, {Babbedge}, {Smith}, {Surace}, {Shupe}, {Fang}, {Farrah}, {Oliver}, {Gonzalez-Solares}, \& {Serjeant}}]{Hatziminaoglou_2004}
{Hatziminaoglou}, E., {Perez-Fournon}, I., {Polletta}, M., {et~al.} 2004, arXiv e-prints, astro

\bibitem[{{Heintz} {et~al.}(2018){Heintz}, {Fynbo}, {H{\o}g}, {M{\o}ller}, {Krogager}, {Geier}, {Jakobsson}, \& {Christensen}}]{Heintz_2018}
{Heintz}, K.~E., {Fynbo}, J.~P.~U., {H{\o}g}, E., {et~al.} 2018, \aap, 615, L8

\bibitem[{{Hewett} {et~al.}(2006){Hewett}, {Warren}, {Leggett}, \& {Hodgkin}}]{Hewett_2006}
{Hewett}, P.~C., {Warren}, S.~J., {Leggett}, S.~K., \& {Hodgkin}, S.~T. 2006, \mnras, 367, 454

\bibitem[{{Ighina} {et~al.}(2023){Ighina}, {Caccianiga}, {Moretti}, {Belladitta}, {Broderick}, {Drouart}, {Leung}, \& {Seymour}}]{Ighina_2023}
{Ighina}, L., {Caccianiga}, A., {Moretti}, A., {et~al.} 2023, \mnras, 519, 2060

\bibitem[{{Ighina} {et~al.}(2025){Ighina}, {Caccianiga}, {Moretti}, {Broderick}, {Leung}, {Rigamonti}, {Seymour}, {Afonso}, {Connor}, {Vignali}, {Wang}, {An}, {Arsioli}, {Bisogni}, {Dallacasa}, {Della Ceca}, {Liu}, {L{\'o}pez-S{\'a}nchez}, {Matute}, {Reynolds}, {Rossi}, {Spingola}, {Severgnini}, \& {Tavecchio}}]{Ighina_2025}
{Ighina}, L., {Caccianiga}, A., {Moretti}, A., {et~al.} 2025, \aap, 698, A158

\bibitem[{{Ivezi{\'c}} {et~al.}(2019){Ivezi{\'c}}, {Kahn}, {Tyson}, {Abel}, {Acosta}, {Allsman}, {Alonso}, {AlSayyad}, {Anderson}, {Andrew}, {Angel}, {Angeli}, {Ansari}, {Antilogus}, {Araujo}, {Armstrong}, {Arndt}, {Astier}, {Aubourg}, {Auza}, {Axelrod}, {Bard}, {Barr}, {Barrau}, {Bartlett}, {Bauer}, {Bauman}, {Baumont}, {Bechtol}, {Bechtol}, {Becker}, {Becla}, {Beldica}, {Bellavia}, {Bianco}, {Biswas}, {Blanc}, {Blazek}, {Blandford}, {Bloom}, {Bogart}, {Bond}, {Booth}, {Borgland}, {Borne}, {Bosch}, {Boutigny}, {Brackett}, {Bradshaw}, {Brandt}, {Brown}, {Bullock}, {Burchat}, {Burke}, {Cagnoli}, {Calabrese}, {Callahan}, {Callen}, {Carlin}, {Carlson}, {Chandrasekharan}, {Charles-Emerson}, {Chesley}, {Cheu}, {Chiang}, {Chiang}, {Chirino}, {Chow}, {Ciardi}, {Claver}, {Cohen-Tanugi}, {Cockrum}, {Coles}, {Connolly}, {Cook}, {Cooray}, {Covey}, {Cribbs}, {Cui}, {Cutri}, {Daly}, {Daniel}, {Daruich}, {Daubard}, {Daues}, {Dawson}, {Delgado}, {Dellapenna}, {de Peyster}, {de Val-Borro}, {Digel}, {Doherty}, {Dubois},
  {Dubois-Felsmann}, {Durech}, {Economou}, {Eifler}, {Eracleous}, {Emmons}, {Fausti Neto}, {Ferguson}, {Figueroa}, {Fisher-Levine}, {Focke}, {Foss}, {Frank}, {Freemon}, {Gangler}, {Gawiser}, {Geary}, {Gee}, {Geha}, {Gessner}, {Gibson}, {Gilmore}, {Glanzman}, {Glick}, {Goldina}, {Goldstein}, {Goodenow}, {Graham}, {Gressler}, {Gris}, {Guy}, {Guyonnet}, {Haller}, {Harris}, {Hascall}, {Haupt}, {Hernandez}, {Herrmann}, {Hileman}, {Hoblitt}, {Hodgson}, {Hogan}, {Howard}, {Huang}, {Huffer}, {Ingraham}, {Innes}, {Jacoby}, {Jain}, {Jammes}, {Jee}, {Jenness}, {Jernigan}, {Jevremovi{\'c}}, {Johns}, {Johnson}, {Johnson}, {Jones}, {Juramy-Gilles}, {Juri{\'c}}, {Kalirai}, {Kallivayalil}, {Kalmbach}, {Kantor}, {Karst}, {Kasliwal}, {Kelly}, {Kessler}, {Kinnison}, {Kirkby}, {Knox}, {Kotov}, {Krabbendam}, {Krughoff}, {Kub{\'a}nek}, {Kuczewski}, {Kulkarni}, {Ku}, {Kurita}, {Lage}, {Lambert}, {Lange}, {Langton}, {Le Guillou}, {Levine}, {Liang}, {Lim}, {Lintott}, {Long}, {Lopez}, {Lotz}, {Lupton}, {Lust}, {MacArthur}, {Mahabal},
  {Mandelbaum}, {Markiewicz}, {Marsh}, {Marshall}, {Marshall}, {May}, {McKercher}, {McQueen}, {Meyers}, {Migliore}, {Miller}, \& {Mills}}]{Ivezic_2019}
{Ivezi{\'c}}, {\v{Z}}., {Kahn}, S.~M., {Tyson}, J.~A., {et~al.} 2019, \apj, 873, 111

\bibitem[{{Iwamoto} {et~al.}(2025){Iwamoto}, {Matsuoka}, {Imanishi}, {Iwasawa}, {Izumi}, {Kashikawa}, {Kawaguchi}, {Sawamura}, {Strauss}, {Takahashi}, \& {Toba}}]{Iwamoto_2025}
{Iwamoto}, R., {Matsuoka}, Y., {Imanishi}, M., {et~al.} 2025, \apj, 979, 183

\bibitem[{{Jiang} {et~al.}(2018){Jiang}, {Hu}, {Xu}, {Dai}, {Zhang}, {Wang}, \& {Chen}}]{Jiang_2018}
{Jiang}, H., {Hu}, Z., {Xu}, M., {et~al.} 2018, in Society of Photo-Optical Instrumentation Engineers (SPIE) Conference Series, Vol. 10702, Ground-based and Airborne Instrumentation for Astronomy VII, ed. C.~J. {Evans}, L.~{Simard}, \& H.~{Takami}, 107022L

\bibitem[{{Jiang} {et~al.}(2016){Jiang}, {McGreer}, {Fan}, {Strauss}, {Ba{\~n}ados}, {Becker}, {Bian}, {Farnsworth}, {Shen}, {Wang}, {Wang}, {Wang}, {White}, {Wu}, {Wu}, {Yang}, \& {Yang}}]{Jiang_2016}
{Jiang}, L., {McGreer}, I.~D., {Fan}, X., {et~al.} 2016, \apj, 833, 222

\bibitem[{{Keller} {et~al.}(2007){Keller}, {Schmidt}, {Bessell}, {Conroy}, {Francis}, {Granlund}, {Kowald}, {Oates}, {Martin-Jones}, {Preston}, {Tisserand}, {Vaccarella}, \& {Waterson}}]{Keller_2007}
{Keller}, S.~C., {Schmidt}, B.~P., {Bessell}, M.~S., {et~al.} 2007, \pasa, 24, 1

\bibitem[{{Kocevski} {et~al.}(2025){Kocevski}, {Finkelstein}, {Barro}, {Taylor}, {Calabr{\`o}}, {Laloux}, {Buchner}, {Trump}, {Leung}, {Yang}, {Dickinson}, {P{\'e}rez-Gonz{\'a}lez}, {Pacucci}, {Inayoshi}, {Somerville}, {McGrath}, {Akins}, {Bagley}, {Bowler}, {Bisigello}, {Carnall}, {Casey}, {Cheng}, {Cleri}, {Costantin}, {Cullen}, {Davis}, {Donnan}, {Dunlop}, {Ellis}, {Ferguson}, {Fujimoto}, {Fontana}, {Giavalisco}, {Grazian}, {Grogin}, {Hathi}, {Hirschmann}, {Huertas-Company}, {Holwerda}, {Illingworth}, {Juneau}, {Kartaltepe}, {Koekemoer}, {Li}, {Lucas}, {Magee}, {Mason}, {McLeod}, {McLure}, {Napolitano}, {Papovich}, {Pirzkal}, {Rodighiero}, {Santini}, {Wilkins}, \& {Yung}}]{Kocevski_2024}
{Kocevski}, D.~D., {Finkelstein}, S.~L., {Barro}, G., {et~al.} 2025, \apj, 986, 126

\bibitem[{{Lambert} {et~al.}(2024){Lambert}, {Assef}, {Mazzucchelli}, {Ba{\~n}ados}, {Aravena}, {Barrientos}, {Gonz{\'a}lez-L{\'o}pez}, {Hu}, {Infante}, {Malhotra}, {Moya-Sierralta}, {Rhoads}, {Valdes}, {Wang}, {Wold}, \& {Zheng}}]{Lambert_2024}
{Lambert}, T.~S., {Assef}, R.~J., {Mazzucchelli}, C., {et~al.} 2024, \aap, 689, A331

\bibitem[{{Lawrence} {et~al.}(2007){Lawrence}, {Warren}, {Almaini}, {Edge}, {Hambly}, {Jameson}, {Lucas}, {Casali}, {Adamson}, {Dye}, {Emerson}, {Foucaud}, {Hewett}, {Hirst}, {Hodgkin}, {Irwin}, {Lodieu}, {McMahon}, {Simpson}, {Smail}, {Mortlock}, \& {Folger}}]{Lawrence_2007}
{Lawrence}, A., {Warren}, S.~J., {Almaini}, O., {et~al.} 2007, \mnras, 379, 1599

\bibitem[{{Lehmer} {et~al.}(2005){Lehmer}, {Brandt}, {Alexander}, {Bauer}, {Schneider}, {Tozzi}, {Bergeron}, {Garmire}, {Giacconi}, {Gilli}, {Hasinger}, {Hornschemeier}, {Koekemoer}, {Mainieri}, {Miyaji}, {Nonino}, {Rosati}, {Silverman}, {Szokoly}, \& {Vignali}}]{Lehmer_2005}
{Lehmer}, B.~D., {Brandt}, W.~N., {Alexander}, D.~M., {et~al.} 2005, \apjs, 161, 21

\bibitem[{{Lodieu} {et~al.}(2014){Lodieu}, {Boudreault}, \& {B{\'e}jar}}]{Lodieu_2014}
{Lodieu}, N., {Boudreault}, S., \& {B{\'e}jar}, V.~J.~S. 2014, \mnras, 445, 3908

\bibitem[{{Mace}(2014)}]{Mace_2014}
{Mace}, G.~N. 2014, PhD thesis, University of California, Los Angeles

\bibitem[{{Madau}(1995)}]{Madau_1995}
{Madau}, P. 1995, \apj, 441, 18

\bibitem[{{Magorrian} {et~al.}(1998){Magorrian}, {Tremaine}, {Richstone}, {Bender}, {Bower}, {Dressler}, {Faber}, {Gebhardt}, {Green}, {Grillmair}, {Kormendy}, \& {Lauer}}]{Magorrian_1998}
{Magorrian}, J., {Tremaine}, S., {Richstone}, D., {et~al.} 1998, \aj, 115, 2285

\bibitem[{{Marchesi} {et~al.}(2016){Marchesi}, {Civano}, {Salvato}, {Shankar}, {Comastri}, {Elvis}, {Lanzuisi}, {Trakhtenbrot}, {Vignali}, {Zamorani}, {Allevato}, {Brusa}, {Fiore}, {Gilli}, {Griffiths}, {Hasinger}, {Miyaji}, {Schawinski}, {Treister}, \& {Urry}}]{Marchesi_2016}
{Marchesi}, S., {Civano}, F., {Salvato}, M., {et~al.} 2016, \apj, 827, 150

\bibitem[{{Marconi} \& {Hunt}(2003)}]{Marconi_2003}
{Marconi}, A. \& {Hunt}, L.~K. 2003, \apjl, 589, L21

\bibitem[{{Marocco} {et~al.}(2021){Marocco}, {Eisenhardt}, {Fowler}, {Kirkpatrick}, {Meisner}, {Schlafly}, {Stanford}, {Garcia}, {Caselden}, {Cushing}, {Cutri}, {Faherty}, {Gelino}, {Gonzalez}, {Jarrett}, {Koontz}, {Mainzer}, {Marchese}, {Mobasher}, {Schlegel}, {Stern}, {Teplitz}, \& {Wright}}]{Marocco_2020}
{Marocco}, F., {Eisenhardt}, P. R.~M., {Fowler}, J.~W., {et~al.} 2021, \apjs, 253, 8

\bibitem[{{Marocco} {et~al.}(2015){Marocco}, {Jones}, {Day-Jones}, {Pinfield}, {Lucas}, {Burningham}, {Zhang}, {Smart}, {Gomes}, \& {Smith}}]{Marocco_2015}
{Marocco}, F., {Jones}, H.~R.~A., {Day-Jones}, A.~C., {et~al.} 2015, \mnras, 449, 3651

\bibitem[{{Matsuoka} {et~al.}(2016){Matsuoka}, {Onoue}, {Kashikawa}, {Iwasawa}, {Strauss}, {Nagao}, {Imanishi}, {Niida}, {Toba}, {Akiyama}, {Asami}, {Bosch}, {Foucaud}, {Furusawa}, {Goto}, {Gunn}, {Harikane}, {Ikeda}, {Kawaguchi}, {Kikuta}, {Komiyama}, {Lupton}, {Minezaki}, {Miyazaki}, {Morokuma}, {Murayama}, {Nishizawa}, {Ono}, {Ouchi}, {Price}, {Sameshima}, {Silverman}, {Sugiyama}, {Tait}, {Takada}, {Takata}, {Tanaka}, {Tang}, \& {Utsumi}}]{Matsuoka_2016}
{Matsuoka}, Y., {Onoue}, M., {Kashikawa}, N., {et~al.} 2016, \apj, 828, 26

\bibitem[{{Matthee} {et~al.}(2024){Matthee}, {Naidu}, {Brammer}, {Chisholm}, {Eilers}, {Goulding}, {Greene}, {Kashino}, {Labbe}, {Lilly}, {Mackenzie}, {Oesch}, {Weibel}, {Wuyts}, {Xiao}, {Bordoloi}, {Bouwens}, {van Dokkum}, {Illingworth}, {Kramarenko}, {Maseda}, {Mason}, {Meyer}, {Nelson}, {Reddy}, {Shivaei}, {Simcoe}, \& {Yue}}]{Matthee_2024}
{Matthee}, J., {Naidu}, R.~P., {Brammer}, G., {et~al.} 2024, \apj, 963, 129

\bibitem[{{Mazzucchelli} {et~al.}(2023){Mazzucchelli}, {Bischetti}, {D'Odorico}, {Feruglio}, {Schindler}, {Onoue}, {Ba{\~n}ados}, {Becker}, {Bian}, {Carniani}, {Decarli}, {Eilers}, {Farina}, {Gallerani}, {Lai}, {Meyer}, {Rojas-Ruiz}, {Satyavolu}, {Venemans}, {Wang}, {Yang}, \& {Zhu}}]{Mazzucchelli_2023}
{Mazzucchelli}, C., {Bischetti}, M., {D'Odorico}, V., {et~al.} 2023, \aap, 676, A71

\bibitem[{{McMahon} {et~al.}(2013){McMahon}, {Banerji}, {Gonzalez}, {Koposov}, {Bejar}, {Lodieu}, {Rebolo}, \& {VHS Collaboration}}]{McMahon+2013}
{McMahon}, R.~G., {Banerji}, M., {Gonzalez}, E., {et~al.} 2013, The Messenger, 154, 35

\bibitem[{{McMahon} {et~al.}(2021){McMahon}, {Banerji}, {Gonzalez}, {Koposov}, {Bejar}, {Lodieu}, {Rebolo}, \& {VHS Collaboration}}]{McMahon_2020}
{McMahon}, R.~G., {Banerji}, M., {Gonzalez}, E., {et~al.} 2021

\bibitem[{{Messias} {et~al.}(2012){Messias}, {Afonso}, {Salvato}, {Mobasher}, \& {Hopkins}}]{Messias_2012}
{Messias}, H., {Afonso}, J., {Salvato}, M., {Mobasher}, B., \& {Hopkins}, A.~M. 2012, \apj, 754, 120

\bibitem[{{Mignoli} {et~al.}(2020){Mignoli}, {Gilli}, {Decarli}, {Vanzella}, {Balmaverde}, {Cappelluti}, {Cassar{\`a}}, {Comastri}, {Cusano}, {Iwasawa}, {Marchesi}, {Prandoni}, {Vignali}, {Vito}, {Zamorani}, {Chiaberge}, \& {Norman}}]{Mignoli_2020}
{Mignoli}, M., {Gilli}, R., {Decarli}, R., {et~al.} 2020, \aap, 642, L1

\bibitem[{{Mortlock} {et~al.}(2012){Mortlock}, {Patel}, {Warren}, {Hewett}, {Venemans}, {McMahon}, \& {Simpson}}]{Mortlock_2012}
{Mortlock}, D.~J., {Patel}, M., {Warren}, S.~J., {et~al.} 2012, \mnras, 419, 390

\bibitem[{{Nakoneczny} {et~al.}(2021){Nakoneczny}, {Bilicki}, {Pollo}, {Asgari}, {Dvornik}, {Erben}, {Giblin}, {Heymans}, {Hildebrandt}, {Kannawadi}, {Kuijken}, {Napolitano}, \& {Valentijn}}]{Nakoneczny_2021}
{Nakoneczny}, S.~J., {Bilicki}, M., {Pollo}, A., {et~al.} 2021, \aap, 649, A81

\bibitem[{{Natarajan} {et~al.}(2024){Natarajan}, {Pacucci}, {Ricarte}, {Bogd{\'a}n}, {Goulding}, \& {Cappelluti}}]{Natarajan_2023}
{Natarajan}, P., {Pacucci}, F., {Ricarte}, A., {et~al.} 2024, \apjl, 960, L1

\bibitem[{{Onken} {et~al.}(2022){Onken}, {Lai}, {Wolf}, {Lucy}, {Hon}, {Tisserand}, {Sokoloski}, {Luna}, {Manick}, {Fan}, \& {Bian}}]{Onken_2022}
{Onken}, C.~A., {Lai}, S., {Wolf}, C., {et~al.} 2022, \pasa, 39, e037

\bibitem[{{Pons} {et~al.}(2019){Pons}, {McMahon}, {Simcoe}, {Banerji}, {Hewett}, \& {Reed}}]{Pons_2019}
{Pons}, E., {McMahon}, R.~G., {Simcoe}, R.~A., {et~al.} 2019, \mnras, 484, 5142

\bibitem[{{Prochaska} {et~al.}(2020){Prochaska}, {Hennawi}, {Westfall}, {Cooke}, {Wang}, {Hsyu}, {Davies}, {Farina}, \& {Pelliccia}}]{Prochaska_2020}
{Prochaska}, J., {Hennawi}, J., {Westfall}, K., {et~al.} 2020, The Journal of Open Source Software, 5, 2308

\bibitem[{{Pudoka} {et~al.}(2024){Pudoka}, {Wang}, {Fan}, {Yang}, {Champagne}, {Jones}, {Bian}, {Cai}, {Jiang}, {Liu}, \& {Wu}}]{Pudoka_2024}
{Pudoka}, M., {Wang}, F., {Fan}, X., {et~al.} 2024, \apj, 968, 118

\bibitem[{{Reed} {et~al.}(2017){Reed}, {McMahon}, {Martini}, {Banerji}, {Auger}, {Hewett}, {Koposov}, {Gibbons}, {Gonzalez-Solares}, {Ostrovski}, {Tie}, {Abdalla}, {Allam}, {Benoit-L{\'e}vy}, {Bertin}, {Brooks}, {Buckley-Geer}, {Burke}, {Carnero Rosell}, {Carrasco Kind}, {Carretero}, {da Costa}, {DePoy}, {Desai}, {Diehl}, {Doel}, {Evrard}, {Finley}, {Flaugher}, {Fosalba}, {Frieman}, {Garc{\'\i}a-Bellido}, {Gaztanaga}, {Goldstein}, {Gruen}, {Gruendl}, {Gutierrez}, {James}, {Kuehn}, {Kuropatkin}, {Lahav}, {Lima}, {Maia}, {Marshall}, {Melchior}, {Miller}, {Miquel}, {Nord}, {Ogando}, {Plazas}, {Romer}, {Sanchez}, {Scarpine}, {Schubnell}, {Sevilla-Noarbe}, {Smith}, {Sobreira}, {Suchyta}, {Swanson}, {Tarle}, {Tucker}, {Walker}, \& {Wester}}]{Reed_2017}
{Reed}, S.~L., {McMahon}, R.~G., {Martini}, P., {et~al.} 2017, \mnras, 468, 4702

\bibitem[{{Richards} {et~al.}(2001){Richards}, {Fan}, {Schneider}, {Vanden Berk}, {Strauss}, {York}, {Anderson}, {Anderson}, {Annis}, {Bahcall}, {Bernardi}, {Briggs}, {Brinkmann}, {Brunner}, {Burles}, {Carey}, {Castander}, {Connolly}, {Crocker}, {Csabai}, {Doi}, {Finkbeiner}, {Friedman}, {Frieman}, {Fukugita}, {Gunn}, {Hindsley}, {Ivezi{\'c}}, {Kent}, {Knapp}, {Lamb}, {Leger}, {Long}, {Loveday}, {Lupton}, {McKay}, {Meiksin}, {Merrelli}, {Munn}, {Newberg}, {Newcomb}, {Nichol}, {Owen}, {Pier}, {Pope}, {Richmond}, {Rockosi}, {Schlegel}, {Siegmund}, {Smee}, {Snir}, {Stoughton}, {Stubbs}, {SubbaRao}, {Szalay}, {Szokoly}, {Tremonti}, {Uomoto}, {Waddell}, {Yanny}, \& {Zheng}}]{Richards_2001}
{Richards}, G.~T., {Fan}, X., {Schneider}, D.~P., {et~al.} 2001, \aj, 121, 2308

\bibitem[{{Richards} {et~al.}(2015){Richards}, {Myers}, {Peters}, {Krawczyk}, {Chase}, {Ross}, {Fan}, {Jiang}, {Lacy}, {McGreer}, {Trump}, \& {Riegel}}]{Richards_2015}
{Richards}, G.~T., {Myers}, A.~D., {Peters}, C.~M., {et~al.} 2015, \apjs, 219, 39

\bibitem[{{Richards} {et~al.}(2004){Richards}, {Nichol}, {Gray}, {Brunner}, {Lupton}, {Vanden Berk}, {Chong}, {Weinstein}, {Schneider}, {Anderson}, {Munn}, {Harris}, {Strauss}, {Fan}, {Gunn}, {Ivezi{\'c}}, {York}, {Brinkmann}, \& {Moore}}]{Richards_2004}
{Richards}, G.~T., {Nichol}, R.~C., {Gray}, A.~G., {et~al.} 2004, \apjs, 155, 257

\bibitem[{{Salvato} {et~al.}(2009){Salvato}, {Hasinger}, {Ilbert}, {Zamorani}, {Brusa}, {Scoville}, {Rau}, {Capak}, {Arnouts}, {Aussel}, {Bolzonella}, {Buongiorno}, {Cappelluti}, {Caputi}, {Civano}, {Cook}, {Elvis}, {Gilli}, {Jahnke}, {Kartaltepe}, {Impey}, {Lamareille}, {Le Floc'h}, {Lilly}, {Mainieri}, {McCarthy}, {McCracken}, {Mignoli}, {Mobasher}, {Murayama}, {Sasaki}, {Sanders}, {Schiminovich}, {Shioya}, {Shopbell}, {Silverman}, {Smol{\v{c}}i{\'c}}, {Surace}, {Taniguchi}, {Thompson}, {Trump}, {Urry}, \& {Zamojski}}]{Salvato_2009}
{Salvato}, M., {Hasinger}, G., {Ilbert}, O., {et~al.} 2009, \apj, 690, 1250

\bibitem[{{Schlegel} {et~al.}(1998){Schlegel}, {Finkbeiner}, \& {Davis}}]{Schlegel_1998}
{Schlegel}, D.~J., {Finkbeiner}, D.~P., \& {Davis}, M. 1998, \apj, 500, 525

\bibitem[{{Secrest} {et~al.}(2015){Secrest}, {Dudik}, {Dorland}, {Zacharias}, {Makarov}, {Fey}, {Frouard}, \& {Finch}}]{Secrest_2015}
{Secrest}, N.~J., {Dudik}, R.~P., {Dorland}, B.~N., {et~al.} 2015, \apjs, 221, 12

\bibitem[{{Selsing} {et~al.}(2016){Selsing}, {Fynbo}, {Christensen}, \& {Krogager}}]{Selsing_2016}
{Selsing}, J., {Fynbo}, J.~P.~U., {Christensen}, L., \& {Krogager}, J.-K. 2016, \aap, 585, A87

\bibitem[{{Shen} {et~al.}(2011){Shen}, {Richards}, {Strauss}, {Hall}, {Schneider}, {Snedden}, {Bizyaev}, {Brewington}, {Malanushenko}, {Malanushenko}, {Oravetz}, {Pan}, \& {Simmons}}]{Shen_2011}
{Shen}, Y., {Richards}, G.~T., {Strauss}, M.~A., {et~al.} 2011, \apjs, 194, 45

\bibitem[{{Stern} {et~al.}(2012){Stern}, {Assef}, {Benford}, {Blain}, {Cutri}, {Dey}, {Eisenhardt}, {Griffith}, {Jarrett}, {Lake}, {Masci}, {Petty}, {Stanford}, {Tsai}, {Wright}, {Yan}, {Harrison}, \& {Madsen}}]{Stern_2012}
{Stern}, D., {Assef}, R.~J., {Benford}, D.~J., {et~al.} 2012, \apj, 753, 30

\bibitem[{{Temple} {et~al.}(2021){Temple}, {Hewett}, \& {Banerji}}]{Temple_2021}
{Temple}, M.~J., {Hewett}, P.~C., \& {Banerji}, M. 2021, \mnras, 508, 737

\bibitem[{{Venemans} {et~al.}(2015){Venemans}, {Ba{\~n}ados}, {Decarli}, {Farina}, {Walter}, {Chambers}, {Fan}, {Rix}, {Schlafly}, {McMahon}, {Simcoe}, {Stern}, {Burgett}, {Draper}, {Flewelling}, {Hodapp}, {Kaiser}, {Magnier}, {Metcalfe}, {Morgan}, {Price}, {Tonry}, {Waters}, {AlSayyad}, {Banerji}, {Chen}, {Gonz{\'a}lez-Solares}, {Greiner}, {Mazzucchelli}, {McGreer}, {Miller}, {Reed}, \& {Sullivan}}]{Venemans_2015}
{Venemans}, B.~P., {Ba{\~n}ados}, E., {Decarli}, R., {et~al.} 2015, \apjl, 801, L11

\bibitem[{{Wang} {et~al.}(2021){Wang}, {Yang}, {Fan}, {Hennawi}, {Barth}, {Banados}, {Bian}, {Boutsia}, {Connor}, {Davies}, {Decarli}, {Eilers}, {Farina}, {Green}, {Jiang}, {Li}, {Mazzucchelli}, {Nanni}, {Schindler}, {Venemans}, {Walter}, {Wu}, \& {Yue}}]{Wang_2021b}
{Wang}, F., {Yang}, J., {Fan}, X., {et~al.} 2021, \apjl, 907, L1

\bibitem[{{Wang} {et~al.}(2016){Wang}, {Elbaz}, {Schreiber}, {Pannella}, {Shu}, {Willner}, {Ashby}, {Huang}, {Fontana}, {Dekel}, {Daddi}, {Ferguson}, {Dunlop}, {Ciesla}, {Koekemoer}, {Giavalisco}, {Boutsia}, {Finkelstein}, {Juneau}, {Barro}, {Koo}, {Micha{\l}owski}, {Orellana}, {Lu}, {Castellano}, {Bourne}, {Buitrago}, {Santini}, {Faber}, {Hathi}, {Lucas}, \& {P{\'e}rez-Gonz{\'a}lez}}]{Wang_2016}
{Wang}, T., {Elbaz}, D., {Schreiber}, C., {et~al.} 2016, \apj, 816, 84

\bibitem[{{Warren} {et~al.}(1987){Warren}, {Hewett}, \& {Osmer}}]{Warren_1987}
{Warren}, S.~J., {Hewett}, P.~C., \& {Osmer}, P.~S. 1987, in Bulletin of the American Astronomical Society, Vol.~19, 1125

\bibitem[{{Wenzl} {et~al.}(2021){Wenzl}, {Schindler}, {Fan}, {Andika}, {Ba{\~n}ados}, {Decarli}, {Jahnke}, {Mazzucchelli}, {Onoue}, {Venemans}, {Walter}, \& {Yang}}]{Wenzl_2021}
{Wenzl}, L., {Schindler}, J.-T., {Fan}, X., {et~al.} 2021, \aj, 162, 72

\bibitem[{{West} {et~al.}(2011){West}, {Morgan}, {Bochanski}, {Andersen}, {Bell}, {Kowalski}, {Davenport}, {Hawley}, {Schmidt}, {Bernat}, {Hilton}, {Muirhead}, {Covey}, {Rojas-Ayala}, {Schlawin}, {Gooding}, {Schluns}, {Dhital}, {Pineda}, \& {Jones}}]{West_2011}
{West}, A.~A., {Morgan}, D.~P., {Bochanski}, J.~J., {et~al.} 2011, \aj, 141, 97

\bibitem[{{Williams} {et~al.}(1996){Williams}, {Blacker}, {Dickinson}, {Dixon}, {Ferguson}, {Fruchter}, {Giavalisco}, {Gilliland}, {Heyer}, {Katsanis}, {Levay}, {Lucas}, {McElroy}, {Petro}, {Postman}, {Adorf}, \& {Hook}}]{Steidel_1996}
{Williams}, R.~E., {Blacker}, B., {Dickinson}, M., {et~al.} 1996, \aj, 112, 1335

\bibitem[{{Wolf} {et~al.}(2020){Wolf}, {Hon}, {Bian}, {Onken}, {Alonzi}, {Bessell}, {Li}, {Schmidt}, \& {Tisserand}}]{Wolf_2020}
{Wolf}, C., {Hon}, W.~J., {Bian}, F., {et~al.} 2020, \mnras, 491, 1970

\bibitem[{{Wright} {et~al.}(2010){Wright}, {Eisenhardt}, {Mainzer}, {Ressler}, {Cutri}, {Jarrett}, {Kirkpatrick}, {Padgett}, {McMillan}, {Skrutskie}, {Stanford}, {Cohen}, {Walker}, {Mather}, {Leisawitz}, {Gautier}, {McLean}, {Benford}, {Lonsdale}, {Blain}, {Mendez}, {Irace}, {Duval}, {Liu}, {Royer}, {Heinrichsen}, {Howard}, {Shannon}, {Kendall}, {Walsh}, {Larsen}, {Cardon}, {Schick}, {Schwalm}, {Abid}, {Fabinsky}, {Naes}, \& {Tsai}}]{Wright_2010}
{Wright}, E.~L., {Eisenhardt}, P. R.~M., {Mainzer}, A.~K., {et~al.} 2010, \aj, 140, 1868

\bibitem[{{Yang} {et~al.}(2024){Yang}, {Schindler}, {Nanni}, {Hennawi}, {Ba{\~n}ados}, {Fan}, {Gloudemans}, {Mazzucchelli}, {Rottgering}, {Venemans}, {Wang}, \& {Yang}}]{Yang_2024}
{Yang}, D.-M., {Schindler}, J.-T., {Nanni}, R., {et~al.} 2024, \mnras, 528, 2679

\bibitem[{{Yang} {et~al.}(2023){Yang}, {Fan}, {Gupta}, {Myers}, {Palanque-Delabrouille}, {Wang}, {Y{\`e}che}, {Aguilar}, {Ahlen}, {Alexander}, {Brooks}, {Dawson}, {de la Macorra}, {Dey}, {Dhungana}, {Fanning}, {Font-Ribera}, {Gontcho}, {Guy}, {Honscheid}, {Juneau}, {Kisner}, {Kremin}, {Le Guillou}, {Levi}, {Magneville}, {Martini}, {Meisner}, {Miquel}, {Moustakas}, {Nie}, {Percival}, {Poppett}, {Prada}, {Schlafly}, {Tarl{\'e}}, {Vargas Magana}, {Weaver}, {Wechsler}, {Zhou}, {Zhou}, \& {Zou}}]{Yang_2023}
{Yang}, J., {Fan}, X., {Gupta}, A., {et~al.} 2023, \apjs, 269, 27

\bibitem[{{Yang} {et~al.}(2020){Yang}, {Wang}, {Fan}, {Hennawi}, {Davies}, {Yue}, {Banados}, {Wu}, {Venemans}, {Barth}, {Bian}, {Boutsia}, {Decarli}, {Farina}, {Green}, {Jiang}, {Li}, {Mazzucchelli}, \& {Walter}}]{Yang_2020a}
{Yang}, J., {Wang}, F., {Fan}, X., {et~al.} 2020, \apjl, 897, L14

\bibitem[{{York} {et~al.}(2000){York}, {Adelman}, {Anderson}, {Anderson}, {Annis}, {Bahcall}, {Bakken}, {Barkhouser}, {Bastian}, {Berman}, {Boroski}, {Bracker}, {Briegel}, {Briggs}, {Brinkmann}, {Brunner}, {Burles}, {Carey}, {Carr}, {Castander}, {Chen}, {Colestock}, {Connolly}, {Crocker}, {Csabai}, {Czarapata}, {Davis}, {Doi}, {Dombeck}, {Eisenstein}, {Ellman}, {Elms}, {Evans}, {Fan}, {Federwitz}, {Fiscelli}, {Friedman}, {Frieman}, {Fukugita}, {Gillespie}, {Gunn}, {Gurbani}, {de Haas}, {Haldeman}, {Harris}, {Hayes}, {Heckman}, {Hennessy}, {Hindsley}, {Holm}, {Holmgren}, {Huang}, {Hull}, {Husby}, {Ichikawa}, {Ichikawa}, {Ivezi{\'c}}, {Kent}, {Kim}, {Kinney}, {Klaene}, {Kleinman}, {Kleinman}, {Knapp}, {Korienek}, {Kron}, {Kunszt}, {Lamb}, {Lee}, {Leger}, {Limmongkol}, {Lindenmeyer}, {Long}, {Loomis}, {Loveday}, {Lucinio}, {Lupton}, {MacKinnon}, {Mannery}, {Mantsch}, {Margon}, {McGehee}, {McKay}, {Meiksin}, {Merelli}, {Monet}, {Munn}, {Narayanan}, {Nash}, {Neilsen}, {Neswold}, {Newberg}, {Nichol}, {Nicinski},
  {Nonino}, {Okada}, {Okamura}, {Ostriker}, {Owen}, {Pauls}, {Peoples}, {Peterson}, {Petravick}, {Pier}, {Pope}, {Pordes}, {Prosapio}, {Rechenmacher}, {Quinn}, {Richards}, {Richmond}, {Rivetta}, {Rockosi}, {Ruthmansdorfer}, {Sandford}, {Schlegel}, {Schneider}, {Sekiguchi}, {Sergey}, {Shimasaku}, {Siegmund}, {Smee}, {Smith}, {Snedden}, {Stone}, {Stoughton}, {Strauss}, {Stubbs}, {SubbaRao}, {Szalay}, {Szapudi}, {Szokoly}, {Thakar}, {Tremonti}, {Tucker}, {Uomoto}, {Vanden Berk}, {Vogeley}, {Waddell}, {Wang}, {Watanabe}, {Weinberg}, {Yanny}, {Yasuda}, \& {SDSS Collaboration}}]{York_2000}
{York}, D.~G., {Adelman}, J., {Anderson}, Jr., J.~E., {et~al.} 2000, \aj, 120, 1579

\end{thebibliography}
\end{document}